\documentclass[prd,%
 preprint,%
groupedaddress,showpacs,nofootinbib,groupedaddress,showpacs,nofootinbib, %
eqsecnum,amsfonts,amsmath,amssymb ]{revtex4} \usepackage{bm}
\usepackage{graphicx}

\def\hideprivate{
   \global\long\def\private##1{
   }
}

\hideprivate    %% This is the default

\begin{document}

\title{Post-Newtonian Freely Specifiable Initial Data for Binary Black Holes in Numerical Relativity}

\date{\today}

\author{Samaya Nissanke}\email{nissanke@iap.fr}
\affiliation{${\mathcal{G}}{\mathbb{R}}\varepsilon{\mathbb{C}}{\mathcal{O}}$
-- Gravitation et Cosmologie,\\ Institut d'Astrophysique de Paris, C.N.R.S.,\\
98$^{\text{bis}}$ boulevard Arago, 75014 Paris, France}

\date{\today}
\begin{abstract}

Construction of astrophysically realistic initial data remains a
central problem when modelling the merger and eventual coalescence
of binary black holes in numerical relativity. The objective of
this paper is to provide astrophysically realistic freely
specifiable initial data for binary black hole systems in
numerical relativity, which are in agreement with post-Newtonian
results. Following the approach taken by Blanchet \cite{B03iv}, we
propose a particular solution to the time-asymmetric constraint
equations, which represent a system of two moving black holes, in
the form of the standard conformal decomposition of the spatial
metric and the extrinsic curvature. The solution for the spatial
metric is given in symmetric tracefree form, as well as in Dirac
coordinates. We show that the solution differs from the usual
post-Newtonian metric up to the 2PN order by a coordinate
transformation. In addition, the solutions, defined at every point
of space, differ at second post-Newtonian order from the exact,
conformally flat, Bowen-York solution of the constraints.

\end{abstract}

\maketitle

\section{Introduction}
\label{Introduction}

\subsection{General Context}

Construction of `astrophysically motivated' initial data is of
utmost importance in efforts directed towards simulating binary
black hole (BBH) mergers. Such simulations are crucial for the
detection and characterization of gravitational waves by the
currently operational ground-based (e.g. LIGO, VIRGO, TAMA,
GEO600), future advanced ground-based and space-based (e.g. LISA)
laser interferometers. The sensitivity of the detectors is
considerably increased by matched filtering of the observed signal
with sets of the expected full waveform. At present, the entire
inspiral and eventual coalescence of the two bodies is viewed in
terms of three consecutive phases. The first and longest duration
phase, the gradual `adiabatic' inspiral, is modelled accurately by
post-Newtonian (PN) methods and solved analytically to an accuracy
of 3.5PN\footnote{The notation 1PN corresponds to the formal $\sim
1/c^2 $ level in a post-Newtonian expansion with respect to the
Newtonian acceleration and gravitational flux (where $ c $ is the
speed of light).} (in the equations of motion and gravitational
flux respectively) \cite{Bliving,BDE04,BDEI04,BI04mult,BDEI04dr}.
When the PN approximation is assumed to be no longer valid,
numerical relativity simulations are used at present to model the
second phase of the plunge and coalescence of the two bodies
\cite{Brueg99,Bra00,BakBrCaLoTa01,BrTiJan04,Al04a,Pre05,Pre05b,CLMZ05,BakCeChKovM05,LSKOR05,HSL06,CLZ06,BakCeChKovM06a,Pre06a}.
After the two compact bodies have merged, the resulting single
object enters the final ringdown phase \cite{PrPullin94}, modelled
accurately using linear perturbation techniques of single black
holes.

Recently, significant advances have been made in the numerical
simulation of the strong-field merger phase
\cite{Pre05b,CLMZ05,BakCeChKovM05,HSL06,CLZ06,BakCeChKovM06a,Pre06a},
in particular, in its dynamical evolution. The latter enables the
accurate description of the BBH's dynamical evolution and the
extraction of an emitted gravitational wave signal. In this
respect, `astrophysically realistic' initial data with an
appropriate gravitational wave content would help to ensure that
the final extracted signal is as close as desirable to the emitted
waveform. Note that the term `astrophysically motivated, or
realistic, data' refers to initial data sets that are constructed
from spacetimes representing astrophysical BBHs, the merger of
which current and future generation interferometers will be able
to detect after several centuries or millennia of their inspiral.
As noted in Section~\ref{related-work}, initial data sets used in
early works have not taken into account the information about the
inspiral and radiation content of BBH systems prior to merger.

Initial data is given traditionally in the standard 3+1 formalism
\cite{ADM62} and its construction is non-trivial. The problem lies
in choosing physically meaningful initial data, amongst an
infinitely large number of non-physical data choices, representing
the BBHs. The most widely used methods to construct initial data,
such as the Extrinsic Curvature (EC) \cite{PfY03} and Conformal
Thin-Sandwich (CTS) \cite{Y99}, adopt conformal decompositions of
both the spatial 3-metric and the extrinsic curvature. This
results in a set of coupled elliptic partial differential
equations, which are to be solved numerically under appropriate
inner and outer boundary conditions, together with a set of freely
specifiable parameters, referred to later in this paper as `free
data'.

\subsection{Related Work}
\label{related-work}

For numerical convenience, early works \cite{Cook94,BrBrug97}
devoted to constructing initial data sets for BBHs introduce a
number of significant simplifying assumptions. These include
conformal flatness, maximal slicing, together with the Bowen-York
extrinsic curvature\footnote{This is the analytical solution to
the decoupled momentum constraint for a conformally flat spacetime
and maximal hypersurface, was first given in \cite{BowY80}, where
it was checked by two surface integrals over a sphere at
infinity.} \cite{BowY80}. However, with considerable recent
advances in spectral methods, as exemplified by
\cite{GGB1,GGB2,PfKST03}, and finite element and finite difference
techniques, limitations of numerical approaches are no longer seen
as a serious obstacle. Moreover, several highly-developed 3D
numerical software
\cite{Pre05b,CLMZ05,BakCeChKovM05,HSL06,CLZ06,BakCeChKovM06a,Pre06a}
for evolving black hole binaries already exist, enabling
gravitational wave extraction. This makes it timely now to go
beyond the initial assumptions of conformal flatness, maximal
slicing and the use of the Bowen-York extrinsic curvature, and to
provide astrophysically motivated initial data.

Among the more recent works that aim to overcome the limitations
of the above assumptions, \cite{MHS99,BoMaNM03,MaM00} use the
alternative double superposed Kerr-Schild approach,
\cite{TiBrCaDi03} uses the PN method in ADM Transverse Traceless
(ADMTT) coordinates, \cite{Alvi00,YTOB05,YT06} use an asymptotic
matching scheme, and \cite{GGB1,GGB2,Cook02,CookP04} develop
helical-Killing vector/quasi-equilibrium initial data. These four
approaches are briefly discussed below; the reader is referred to
\cite{cookliving} and \cite{Pf04} for a more detailed discussion.

Taking the double superposed Kerr-Schild approach first, proposed
initially in \cite{MHS99} and developed further in
\cite{BoMaNM03,MaM00}, it is important to note that Kerr
spacetimes do not admit conformally flat slices. In this approach,
the free data are constructed from a linear superposition of two
single (spinning or boosted) black holes in Kerr-Schild
coordinates representing the spatial conformal metric.
Furthermore, the Kerr-Shild data advantageously reduce into
stationary Kerr in the specific limit of two widely-separated
black holes. The Kerr-Schild free data have also been numerically
implemented \cite{BoMaNM03} to obtain full initial data
sets\footnote{These represent the solution, obtained numerically,
to the coupled elliptic partial differential constraint equations
using the freely specifiable parameters, or free data.} and are
amongst the more extensively studied of all astrophysically
motivated free data; see, for example, \cite{PfCookT02}. However,
as this work notes (see Appendix~\ref{Kerr-Schild-initial-data}),
the superposed Kerr-Schild initial data disagrees with
post-Newtonian results at the second-post-Newtonian order.

In relation to PN free data, similar to Kerr spacetimes, binary
black hole systems are also not conformally flat at second
post-Newtonian order \cite{Ri97,DJSisco}. It is widely acknowledged
now that post-Newtonian results modelling the gradual inspiral
should suggest a more astrophysically relevant form of the free data
for the last few stable orbits prior to the Innermost Circular Orbit
(ICO)\footnote{The ICO is defined by the minimum of the binary's
energy function for circular orbits. The definition does not
include, significantly, radiation reaction terms and, therefore, is
not physically meaningful in the context of exact radiative
solutions. It is well-defined, however, for both numerical
\cite{GGB1,GGB2} and 3PN calculations, and allows for comparisons
between the two facilitating greater understanding.}. This is an
important motivation for our work, and has also been considered in
the PN method in ADMTT coordinates \cite{TiBrCaDi03}.

The PN method in ADMTT coordinates due to Tichy {\sl et al.}
\cite{TiBrCaDi03} proposes the direct adoption of the 2PN metric
in ADM coordinates as the conformal spatial metric. The scheme
outlined in \cite{TiBrCaDi03} is for a numerically computed
conformal factor to `correct' the initial post-Newtonian solution
and, thus, to account for higher-order post-Newtonian terms in the
physical metric, which were initially neglected in the proposed
conformal counterpart. The PN initial data in \cite{TiBrCaDi03}
are based on the PN results in the ADMTT gauge. Motivations given
there for the choice of ADMTT gauge apply equally well to PN free
data in the harmonic gauge, which is used in our work. Among these
commonalities between the two methods, i.e. \cite{TiBrCaDi03} and
our work, firstly, both methods easily find expressions for the
conformal 3-metric and extrinsic curvature without any logarithmic
terms. Secondly, to the 1.5PN order $\sim {\cal
O}(\frac{1}{c^3})$, the freely specifiable data sets due to both
methods also agree with the puncture approach \cite{BrBrug97} and
the choice in a specific hypersurface where the trace of the
extrinsic curvature, $K$, (otherwise known as the `mean
curvature') vanishes. Finally, the two free data sets do not
include, for the time being at least, any radiation content.

In the asymptotic matching scheme, following the initial work by
Alvi \cite{Alvi00}, Yunes {\sl et al.} \cite{YTOB05,YT06} propose
an approximate metric for corotating BBHs spacetimes.
Advantageously, this includes both 1PN effects, and, unlike in
post-Newtonian free data sets due to Tichy {\sl et al.}
\cite{TiBrCaDi03} discussed above and our work, tidal effects
experienced by each black hole when close to one another. The
metric is obtained by asymptotically matching a post-Newtonian
metric for a binary system to a perturbed Schwarzschild metric
modelling each black hole. At present, the method must be first
projected onto the constraint hypersurface before numerical
implementation. In addition, the method currently employs
post-Newtonian results at the 1PN order, and considers the
specific configuration of corotating BBHs.

Turning our attention to the fourth approach, that is,
construction of helical-Killing vector/quasi-equilibrium initial
data, Gourgoulhon {\sl et al.} \cite{GGB1,GGB2} and Cook
\cite{Cook02} first showed, by assuming an approximate
helical-Killing vector for orbits prior to merger, a conformally
flat spacetime and maximal slicing, that the four constraints plus
one evolution equation simplify considerably in the CTS
decomposition. The resulting set of equations is solvable
numerically using spectral methods. Subtle problems, however,
remain with the choice of inversion-symmetric inner boundary
conditions at excision spheres around the singularities of the
black holes - an implicit consequence of specifying a Misner
topology. In this respect, Cook {\sl et al.} \cite{CookP04}
recently proposed other physically-motivated quasi-equilibrium
boundary conditions, which allow for arbitrary specifications of
the conformal metric, mean curvature, and the shape of the
excision regions. As is mentioned later, helical-Killing
vector/quasi-equilibrium initial data sets have been studied in
detail through construction of quasi-circular orbits and in
comparison with post-Newtonian work; see \cite{YoCoShBau04,Ha05}
and \cite{DGG,CookP04} respectively. An advantage over the double
superposed Kerr-Schild method is that this approach does not
require the specification of the extrinsic curvature. In addition,
BBH dynamical evolutions have been performed recently for
quasi-circular orbits. For example, \cite{Al04a} uses a scheme
based on excision, new gauges, the BSSN formulation
\cite{NaOKo87,ShNa95,BaS99,Aletal00}, and a corotating frame,
together with initial data constructed by means of the puncture
method \cite{BrBrug97}, whereas \cite{Pre06b} uses the generalized
harmonic evolution scheme \cite{Pre05,Pre05b,Pre06a} based on
Cook-Pfeiffer initial data \cite{CookP04}.

In contrast to the above methods, Blanchet \cite{B03iv} presents
an alternative set of PN motivated free data for the
time-symmetric constraints, corresponding to two Schwarzschild
black holes momentarily at rest\footnote{The exact solutions of
Misner \cite{Misner63}, Lindquist \cite{Lind63} and Brill and
Lindquist \cite{BrillLind63} to the time-symmetric constraints for
conformally flat spacetimes provides an insight into the
geometrostatic nature of two black holes, and in particular, into
their topology: the solution for multiple black holes was directly
inspired from the case of a single Schwarzschild black hole
containing a Einstein-Rosen bridge (alternatively known as a
`worm-hole') joining two asymptotically flat universes. The Misner
solution refers to two black holes, each containing an
Einstein-Rosen bridge, which connects our universe to a second
asymptotically flat region. In contrast, the Brill-Lindquist joins
our universe to two distinct separate universes through two
Einstein-Rosen bridges. Furthermore, it allows the computation at
infinity of both the total ADM mass of the binary and each
separate black hole, which then determines the gravitational
binding energy of the system in the center of mass frame.}. This
work proposes a particular solution to the time-symmetric
constraints in the form of the conformal metric,
$\tilde{\gamma}_{ij}$. The solution is such that the
post-Newtonian expansion of the corresponding physical metric,
$\gamma_{ij}$, is isometric to the standard post-Newtonian metric
in harmonic coordinates at 2PN order as $c \rightarrow + \infty$.
Furthermore, the solution is defined globally in space and differs
from the Brill-Lindquist solution \cite{BrillLind63} for
conformally-flat spacetimes at 2PN. It is important to note that
for the numerical implementations of BBH coalescence, the
time-symmetric constraints involved relate to physically
unrealistic events. This is because, from a general astrophysical
perspective, black hole binaries prior to merger possess a
velocity and, hence, impart a non-zero value to the extrinsic
curvature to the hypersurface, i.e. $K_{ij} \neq 0 $, $K_{ij}$
being the physical extrinsic curvature.

In this respect, our work aims to establish physically realistic
free data for the plunge and merger phase of two black holes in
numerical relativity, in conformity with post-Newtonian results
\cite{BFP98}. It follows from, and develops further, the
time-symmetric approach proposed in \cite{B03iv}.

Before entering into the details of our work, we remark on some of
the difficulties in assessing how significant is the deviation
arising from the assumption of a conformally flat metric for the
merger of BBHs. Firstly, as shown by Pfeiffer {\sl et al.}
\cite{PfCookT02}, the resulting initial data sets are particularly
sensitive to the choice in extrinsic curvature, as opposed to in
the conformal 3-metric, $\tilde{\gamma}_{ij}$. Their work is based
on a comparison of initial data sets based on three different
conformal decompositions of the initial constraints for BBHs.
These initial data sets have been constructed by using both the
double superposition of two black holes
\cite{MHS99,MaM00,BoMaNM03} and the Bowen-York solution
\cite{BowY80}. Note that although the superposed Kerr-Schild data
sets go beyond the standard assumption of conformal flatness, they
disagree with post-Newtonian results at 2PN order (see
Appendix~\ref{Kerr-Schild-initial-data} of this paper). Secondly,
as is mentioned earlier, the recent results, in particular, the
location of the ICO based on helical-Killing
vector/quasi-equilibrium initial data
\cite{GGB1,GGB2,Cook02,CookP04} for co-rotating and irrotational
BBHs, show strong agreement with PN predictions
\cite{DGG,CookP04}. The agreement with PN results is not
surprising, considering that helical symmetry is exact with
respect to 2PN order. Despite encouraging agreement with
post-Newtonian predictions, all the currently used
quasi-equilibrium initial data sets assume `physically improbable'
conformally flat spacetimes.

\subsection{An Overview of Our Contribution}
\label{contribution}

In this work, we propose a particular solution in the form of
conformal spatial metric, $\tilde{\gamma}_{ij}$, and conformal
extrinsic curvature, $\tilde{K}_{ij}$, to the time-asymmetric
constraint equations that represent a system of two moving black
holes. The solution is chosen such that the post-Newtonian
re-expansions of the corresponding physical metric, $\gamma_{ij}$,
and physical extrinsic curvature, $K_{ij}$, are both isometric to
the post-Newtonian metric in harmonic coordinates and
post-Newtonian extrinsic curvature up to the 2PN order. Note that
the post-Newtonian metric is formally valid in the source's ``near
zone''\footnote{The size of the near zone is much smaller than the
characteristic gravitational wavelength.}. Importantly for our
work, it has also been proved that the post-Newtonian metric
arises itself from the re-expansion of a ``global''
post-Minkowskian (expansion in G) multipolar solution as $c
\rightarrow \infty$, defined everywhere in spacetime, including
the wave zone \cite{PB02}. The solution for the conformal metric,
$\tilde{\gamma}_{ij}$, is given, firstly, in a symmetric
trace-free form, similar to the approach instigated in
\cite{B03iv} and, secondly, for the first time, in Dirac
coordinates. The motivation behind the choice of Dirac coordinates
is to provide free data for constrained schema of the Einstein
equations, such as the one based on a covariant generalized Dirac
gauge\footnote{\private{The covariant generalized Dirac gauge,
defined as (\ref{InitialConstraint2EC}) in \cite{BGGN04}, is a
development of the gauge first introduced by Dirac in 1959 in
order to fix the coordinates in a Hamiltonian formulation of
general relativity, prior to its quantization. The new generalized
variant resolves the issue of its incovariance under coordinate
transformation, $x^i \rightarrow x'^i$ on the hypersurface
$\Sigma_t$. This was raised by Smarr and York \cite{SmarrYork} in
their initial discussion for using the original definition as a
radiation gauge in numerical relativity.} As discussed in
\cite{BGGN04}, the advantages of implementing the covariant
generalized Dirac gauge, defined as (\ref{InitialConstraint2EC})
in \cite{BGGN04}, are numerous. Firstly, it fully specifies the
coordinates in the slice $\Sigma_t$ (up to some inner boundary
conditions for a slice containing holes); the latter property
allows for the search for stationary solutions for the proposed
set of equations, for instance, quasi--stationary initial
conditions. In addition, the choice in gauge also results
asymptotically in transverse--traceless (TT) coordinates, which
are attractive for treating the problem of gravitational radiation
and are analogous to the Coulomb gauge in electromagnetism. In
addition, we show here that by relating our physical metric in
Dirac coordinates, $\gamma_{ij}^{\sl{Dirac}}$, to the
post-Newtonian metric, the resulting conformal factor
$\Psi^{\mathrm{DIRAC}}$ assumes a simple form at 2PN order; see
eqn.~(\ref{psiDIRAC}) in
Section~\ref{Relationship-between-conformal-metrics}.} and
spherical coordinates, as proposed recently by Bonazzola {\sl et
al.} \cite{BGGN04}.

Furthermore, our PN free data is constructed in such a way that it
is defined at every point in space and differs from the Bowen-York
solution at 2PN order. By demonstrating equivalence relationships
between our PN free data and the full 2PN metric, we also prove that
the Bowen-York extrinsic curvature is physically equivalent to the
2PN derived extrinsic curvature. In detail, we show that each
extrinsic curvature is defined with respect to a different
hypersurface corresponding to a distinct 3+1 foliation of the full
spacetime.

It is important to note that this work proposes only the freely
specifiable parameters for initial data, hence, providing a free
data scheme. The full physical initial data set is only possible
after the constraint equations have been solved numerically, with
appropriate inner and outer boundary conditions. Present numerical
methods for generating a `constraint-satisfying' data set from
some physically motivated freely specifiable parameters include
solving the constraints either using second-order finite
difference techniques together with multigrid \cite{TiBrCaDi03} or
successive over-relaxation \cite{BoMaNM03}, or using spectral
methods \cite{GGB1,PfCookT02}. Finite regions very close to the
black hole singularities are frequently removed from the
computational domain using standard excision techniques.

\subsection{Structure of the Paper}

The structure of the paper is as follows.
Section~\ref{Conformal-decomposition} introduces the constraint
equations in the 3+1 formalism required for presenting our
approach and results. We outline the two standard conformal
decompositions of the constraints: the Extrinsic Curvature (EC)
and Conformal Thin Sandwich (CTS) methods, corresponding to a
Hamiltonian or Lagrangian viewpoint of the problem respectively.
In Section~\ref{PN-free-data}, we first present the proposed PN
free data in both EC and CTS decompositions, namely $(
\tilde{\gamma}_{ij}, \tilde{A}_{\mathrm{TT}}^{ij}, K,
\tilde{\sigma})$ and $(\tilde{\gamma}_{ij}, \tilde{u}^{ij}, K,
\tilde{N}\ \mathrm{or}\
\partial_t K)$, respectively. Section~\ref{Proof-of-equivalence} subsequently outlines the
proof of equivalence between our PN initial data and the full 2PN
metric in harmonic coordinates, thus justifying the results given
in Section~\ref{PN-free-data}. We then briefly investigate the
global properties of our proposed `near-zone' PN solution and show
that it is defined at every point in space. Based on our proofs,
we finally show that the Bowen-York extrinsic curvature is
physically equivalent to the 2PN derived extrinsic curvature.
Section~\ref{conclusion} concludes the paper with a summary of
results.

\section{3+1 Conformal Decomposition}
\label{Conformal-decomposition}

The purpose of this section is to review the basic concepts
related to the Extrinsic Curvature and Conformal Thin Sandwich
decompositions of the constraints; this is necessary for the
presentation of the results on PN free data in
Section~\ref{PN-free-data}. Our account here follows that in
\cite{cookliving,Pf04}, where more details may be found.

\subsection{Preliminaries}
\label{Preliminaries}

In the Cauchy formalism of Einstein's equations, a globally
hyperbolic space-time is foliated into a set of space-like 3
dimensional hypersurfaces, $\Sigma_t$, where each slice is a
constant hypersurface parameterized by the time coordinate $t$ ($
t \in \mathbb{R} $). According to the standard 3+1 decomposition,
$n^{\mu}$, a future-directed time-like unit normal to $\Sigma_t$,
is defined as\private{\begin{equation}
\label{futuretimelikeunitnormal}}$ n^{\mu} \equiv - N \nabla^{\mu}
t $, \private{\end{equation}}where $N$ denotes the {\sl lapse
function}\footnote{The {\sl lapse antidensity}, $\alpha$, a scalar
of weight -1, is the undetermined multiplier of the scalar
Hamiltonian constraint, which is in agreement with the canonical
form of the ADM action. The relation between $\alpha$ and $N$ is
discussed later.}. Together with the causal stability of globally
hyperbolic spacetimes, the global time vector field, $t^{\mu}$, is
given by $t^{\mu} = N n^{\mu} + \beta^{\mu}$, which is both
transverse to $\Sigma_t$, and is normalized such that $t^{\alpha}
\nabla_{\alpha} t = 1$. The {\sl shift vector}, $\beta^{\mu}$,
thus trivially fulfills the condition $\beta^{\alpha} n_{\alpha} =
0$. Following the definition of $n^{\mu}$, the spacetime metric
$g_{\mu \nu}$ induces a purely spatial metric, $\gamma_{ij}$,
defined as $\gamma_{\mu \nu} = g_{\mu \nu} + n_{\mu} n_{\nu}$,
\private{\begin{equation} \label{gammaij} \gamma_{\mu \nu} =
g_{\mu \nu} + n_{\mu} n_{\nu}. \end{equation}}where subscripts in
Latin and Greek characters denote three-dimensional and
four-dimensional indices respectively.

The spacetime metric components, $g_{\mu \nu}$, are thus expressed
with respect to a preferred coordinate system $(t, x^{i})$ as,
\begin{equation}
\label{metricinterval} g_{\mu \nu} dx^{\mu} dx^{\nu}  \equiv ds^2
 =  - N^2 (dt)^2 + \gamma_{ij} (dx^i + \beta^i dt)(dx^j +
\beta^j dt).
\end{equation}
Together with $\gamma_{ij}$, a well-posed initial value problem must
also specify how each spacelike hypersurface is embedded into the
full spacetime. This is achieved by introducing the extrinsic
curvature tensor, $K_{ij}$, i.e.,
\begin{eqnarray}
\label{Kmunudefn} \nonumber K_{\mu \nu} & \equiv & -\frac{1}{2}
\gamma^{\delta}_{\mu} \gamma^{\rho}_{\nu} {\cal L}_n
\gamma_{\delta \rho} \\ \label{Kijdefn3} & \equiv & -
\gamma^{\rho}_{\mu} \nabla_{\rho} n_{\nu}
\end{eqnarray}
or equivalently,
\begin{equation}
\label{Kijdefn}
K_{ij}  \equiv -  \frac{1}{2} {\cal L}_n \gamma_{ij},
\end{equation}
where ${\cal L}_n$ represents the Lie derivative along the normal $n^{\mu}$
direction. By using the Gauss-Codazzi relations, the Einstein field equations
are expressed in terms of six evolution equations,
\begin{eqnarray}
\label{evolution1} \nonumber \partial_t K_{ij}  & = & N  [
\bar{R}_{ij} - 2 K_{il} K^l_j + K K_{ij} - 8 \pi G S_{ij} + 4 \pi
G \gamma_{ij} (S - \rho)  ] -  \bar{\nabla}_i \bar{\nabla}_j N
\\ \label{evolution1} & & +  \beta^l \bar{\nabla}_l K_{ij}  +
 K_{il} \bar{\nabla}_j \beta^l  +  K_{jl} \bar{\nabla}_i
\beta^l,
\end{eqnarray}
and four initial constraint equations,
\begin{eqnarray}
\label{constraint1}
\bar{R} + K^2 - K_{ij} K^{ij} & = & 16 \pi G \rho, \\
\label{constraint2} \bar{\nabla}_j ( K^{ij} - \gamma^{ij} K ) & =
& 8 \pi G j^i,
\end{eqnarray}
where eqns.~(\ref{constraint1}) and (\ref{constraint2}) are referred
to as the {\sl Hamiltonian} and {\sl momentum} constraints
respectively. In (\ref{evolution1})-(\ref{constraint2}), $\rho$
denotes the matter energy density; $S_{ij}$ the matter stress tensor
with $S \equiv S^i_i$, and $j^i$ the matter momentum density. In
addition, $K \equiv K^i_i$ gives the trace of the extrinsic
curvature, also known as the {\sl mean curvature}, and
$\bar{R}_{ij}$ is the 3-Ricci tensor associated with $\gamma_{ij}$.
In order to avoid confusion, we follow the convention in
\cite{cookliving} that covariant derivative and the Ricci tensor
associated with the physical 3-metric $\gamma_{ij}$ are written with
overbars as $\bar{\nabla}_j $ and $\bar{R}_{ij}$.

Definition (\ref{Kijdefn}) additionally results in a kinematic relation
between $K_{ij}$ and $\gamma_{ij}$,
\begin{equation}
\label{kinematic1}
\partial_t \gamma_{ij}  =  - 2 N K_{ij}  +  \bar{\nabla}_i
\beta_j  +  \bar{\nabla}_j \beta_i,
\end{equation}
complementing the set of equations
(\ref{evolution1})-(\ref{constraint2}). Theoretically, the
constraint equations are thus preserved exactly under evolution
(due to the presence of the Bianchi identities).

Solutions to the most general form of the initial-value
constraints for an extrinsic curvature $K_{ij} \neq 0 $, as shown
by equations (\ref{constraint1}), (\ref{constraint2}), are
non-trivial, partly because the equations do not fully
specify\footnote{Constraints (\ref{constraint1}) and
(\ref{constraint2}) restrict four out of the twelve degrees of
freedom in $(\gamma_{ij}, K_{ij})$.} the dynamical, constrained or
gauge nature of the components of the 3-metric, $\gamma_{ij}$, and
extrinsic curvature, $K_{ij}$. Moreover, it becomes even harder to
find solutions to the constraints for the specific case of two
black holes. Standard procedures to solve the initial-value
problem adopt conformal decompositions of the metric and of the
extrinsic curvature; see \cite{Y74,L44,ChoquetYork,Wald}.

The physical 3-metric, $\gamma_{ij}$, is assumed to be conformally
equivalent to a non-physical background metric,
$\tilde{\gamma}_{ij}$, by a conformal factor $\Psi$, i.e.,
\begin{equation}
\label{conformalmetricdefn} \gamma_{ij}   =   \Psi^{ 4 }
\tilde{\gamma}_{ij},
\end{equation}
where $\Psi$ is the strictly positive background (conformal)
factor.

The Hamiltonian constraint equation, (\ref{constraint1}), as first
proposed by Lichnerowicz in \cite{L44}, is thus re-expressed for a
vacuum as,
\begin{eqnarray}
\label{constraint3} \tilde{\Delta}   \Psi   -   \frac{ 1 }{ 8 }
\Psi \tilde{R} - \frac{1}{8} \Psi^5 K^2 + \frac{1}{ 8 }  \Psi^{5 }
K_{ij} K^{ij} & = & 0,
\end{eqnarray}
where $\tilde{\Delta} \equiv \tilde{\nabla^i} \tilde{\nabla_i}$ is
the scalar Laplacian operator, and $\widetilde{\nabla}_{j}$ and
$\tilde{R}_{ij}$ are the covariant derivative and Ricci tensor
associated with the conformal spatial metric,
$\tilde{\gamma}_{ij}$ (a tilde distinguishing, here and elsewhere,
all quantities that have a conformal relationship with quantities
in the physical space). A full conformal decomposition, however,
requires introducing a conformal extrinsic curvature,
$\tilde{K}_{ij} $, as developed by York \cite{Y74}. Our work
concerns the two most widely used decompositions, which are seen
to be consistent with each other: a) the fully conformally
covariant EC decomposition, introduced recently in \cite{PfY03}
and which improves on earlier non-conformally covariant
decompositions, i.e. the Conformal Transverse Traceless (CTT) and
Physical Transverse Traceless (PTT), and b) the CTS methods. Both
methods involve expressing $K_{ij}$ in terms of its trace and
trace-free constituents,
\begin{equation}
\label{conformal3} K_{ij}   \equiv   A_{ij}  +  \frac{1}{3}
\gamma_{ij} K.
\end{equation}
Each decomposition differs in its subsequent treatment of the
symmetric trace-free extrinsic curvature, $A_{ij}$. The EC method
is based on the 3-metric, $\gamma_{ij}$, and extrinsic curvature,
$K_{ij}$, where the latter is related uniquely to the canonical
momentum, $\bar{\pi}^{ij}$. It provides a `Hamiltonian' viewpoint
to the conformal decomposition of the constraints; see
\cite{Pf04}. In contrast, the CTS decomposition considers the
conformal metric, $\tilde{\gamma}_{ij}$, and its time derivative,
together with the mean curvature, $K$, and the conformal lapse
function, $\tilde{N}$ (where, as discussed in Section~\ref{CTS},
$\tilde{N}$ is related to the time-derivative of $K$ through a
fifth coupled elliptic equation). It offers, therefore, a
complementary `Lagrangian' approach to the conformal decomposition
of the initial data. The EC and CTS decompositions require a
different set of free data, in particular, with respect to the
extrinsic curvature or its tensor equivalent in the CTS instance.
Both decompositions also result in, as illustrated in
\cite{PfCookT02}, different physical initial data.
Sections~\ref{EC} and \ref{CTS} summarize the main results of the
two decompositions.

\subsection{Extrinsic Curvature Decomposition - {\sl Hamiltonian formalism}}
\label{EC}

The extrinsic curvature method (see \cite{PfY03} for its proposed
formulation and \cite{Pf04} for a detailed review) begins by
applying a weighted transverse traceless decomposition to the
tracefree extrinsic curvature, $A_{ij}$, i.e.,
\begin{equation}
\label{ECAij2} A^{ij}    =   \frac{1}{\sigma}(\bar{\mathbb{L}}
X)^{ij} + A_{\mathrm{TT}}^{ij},
\end{equation}
where $A_{\mathrm{TT}}^{ij}$ is a symmetric, tranverse-tracefree
tensor and $\sigma$ is a strictly positive and bounded function on
the 3-dimensional hypersurface, $\Sigma_t$, $0 < \epsilon \leq
\sigma < \infty$ for $\epsilon = constant$. Given the vector field,
$X^i$, the notation $\bar{\mathbb{L}}$ refers to the conformal
longitudinal operator, $(\bar{\mathbb{L}} X)^{ij}  \equiv
\bar{\nabla}^i X^j + \bar{\nabla}^j X^i - \frac{2}{3} \gamma^{ij}
\bar{\nabla}_l X^l$, satisfying $(\bar{\mathbb{L}} X)^{ij}  =
\Psi^{-4} (\tilde{\mathbb{L}} X)^{ij}$.\private{
\begin{equation}
\label{Longoperator} (\bar{\mathbb{L}} X)^{ij}  \equiv
\bar{\nabla}^i X^j + \bar{\nabla}^j X^i - \frac{2}{3} \gamma^{ij}
\bar{\nabla}_l X^l.
\end{equation}
for  and which satisfies,
\begin{equation}
\label{LongoperatorII} (\bar{\mathbb{L}} X)^{ij}  =  \Psi^{-4}
(\tilde{\mathbb{L}} X)^{ij}
\end{equation}} Divergence of the symmetric trace-free tensor,
$A^{ij}$, results in the following expression\footnote{It involves
the well-defined elliptic operator in divergence form,
$\bar{\nabla}_j [\sigma^{-1} (\bar{\mathbb{L}}.)^{ij}]$, as
discussed in \cite{PfY03} and \cite{Pf04}. For $\sigma = 1$, the
latter operator reduces to the vector Laplacian, $(\bar{\Delta}_L
Y)^i \equiv \bar{\nabla}_j (\bar{\mathbb{L}} X)^{ij}$. As discussed
in \cite{PfY03}, this is solvable on compact and on asymptotically
flat manifolds, given certain asymptotic conditions (\cite{Y74}
demonstrates the existence and uniqueness of a solution to
(\ref{ECAij2}) for closed manifolds). In the case of non-compact
manifolds without boundaries, boundary conditions must always be
specified and their choice will directly affect the solution of
$X^i$. As remarked in \cite{PfY03}, there is no uniqueness property
without boundary conditions.}, $\bar{\nabla}_j A^{ij} =
\bar{\nabla}_j [ \frac{1}{\sigma}(\bar{\mathbb{L}} X)^{ij} ]$, which
can be solved for $X^i$.\private{\begin{equation} \label{ECAij2a}
\bar{\nabla}_j A^{ij}    =   \bar{\nabla}_j [
\frac{1}{\sigma}(\bar{\mathbb{L}} X)^{ij} ],
\end{equation}
involving the well-defined elliptic operator in divergence form,
$\bar{\nabla}_j [\sigma^{-1} (\bar{\mathbb{L}}.)^{ij}]$, as
discussed in \cite{PfY03} and \cite{Pf04}. \footnote{For $\sigma =
1$, the latter operator reduces to the vector Laplacian,
$(\bar{\Delta}_L Y)^i \equiv \bar{\nabla}_j (\bar{\mathbb{L}}
X)^{ij}$; as discussed in \cite{PfY03}, this is solvable on
compact and on asymptotically flat manifolds, given certain
asymptotic conditions (\cite{Y74} demonstrates the existence and
uniqueness of a solution to (\ref{ECAij2}) for closed manifolds).
In the case of non-compact manifolds without boundaries, boundary
conditions must always be specified and their choice will directly
affect the solution of $X^i$. As remarked in \cite{PfY03}, there
is no uniqueness property without boundary conditions.}}

Having obtained $A^{ij}_{\mathrm{TT}}$ following the substitution of
$X^i$ in (\ref{ECAij2}), the subsequent conformal scaling of
$A^{ij}_{\mathrm{TT}} \equiv    \Psi^{-10}
\tilde{A}^{ij}_{\mathrm{TT}}$ and $ \sigma \equiv \Psi^{6}
\tilde{\sigma}$ allows for,
\begin{equation} \label{ECAij} A^{ij}    =  \Psi^{-10} \left(
\tilde{A}^{ij}_{\mathrm{TT}} + \frac{1}{\tilde{\sigma}}
(\tilde{\mathbb{L}} X)^{ij} \right)   =  \Psi^{-10} \tilde{A}^{ij},
\end{equation} where the weighted
transverse trace-free decomposition in conformal space is
$\tilde{A}^{ij}    \equiv \tilde{A}^{ij}_{\mathrm{TT}} +
\frac{1}{\tilde{\sigma}} \left( (\tilde{\mathbb{L}} X)^{ij}
\right)$.

Introduction of the weight function $\sigma$ in \cite{PfY03}
ensures that the extrinsic curvature decomposition commutes with
conformal transformations of the free data\footnote{This is in
contrast to its earlier variants (i.e. the CTT and PTT
decompositions), where the conformal transformation and
transverse-traceless decomposition are non-commutative
operations.}. Such a specific choice in the conformal scaling of
$\sigma$ also allows $\sigma$ to be related directly to the lapse
function, $N$. As discussed in Section~\ref{CTS}, the conformal
$\tilde{N}$ is a component of the free data in the CTS
decomposition.

In conjunction with $\bar{\nabla}_j (\Psi^{-10} \tilde{S}^{ij} ) =
\Psi^{-10} \tilde{\nabla}_j \tilde{S}^{ij}$, the complete set of
elliptic constraint equations for a vacuum are,
\begin{eqnarray}
\widetilde{\nabla}_j \left( \frac{1}{\tilde{\sigma}} (
\tilde{\mathbb{L}} X)^{ij} \right) - \frac{2}{3} \Psi^6
\widetilde{\nabla}^i K & = & 0, \label{InitialConstraint1EC} \\
\widetilde{\Delta} \Psi - \frac{1}{8} \Psi \tilde{R} -
\frac{1}{12} \Psi^5 K^2 + \frac{1}{8} \Psi^{-7} \tilde{A}_{ij}
\tilde{A}^{ij} & = & 0. \label{InitialConstraint2EC}
\end{eqnarray}
In this case, the set of free data comprises $(
\tilde{\gamma}_{ij}, \tilde{A}_{\mathrm{TT}}^{ij}, K,
\tilde{\sigma})$, which enables the solution to
(\ref{InitialConstraint1EC}) and (\ref{InitialConstraint2EC}) for
$\Psi$ and $X^i$ with appropriate inner and outer boundary
conditions. Hence, given relationships
(\ref{conformalmetricdefn}), (\ref{conformal3}),
$A^{ij}_{\mathrm{TT}}    \equiv \Psi^{-10}
\tilde{A}^{ij}_{\mathrm{TT}}$, and  $\tilde{A}^{ij} \equiv
\tilde{A}^{ij}_{\mathrm{TT}} + \frac{1}{\tilde{\sigma}} \left(
(\tilde{\mathbb{L}} X)^{ij} \right)$, it is possible to construct
the physical initial data, $\gamma_{ij}$ and $K_{ij}$.

\subsection{Conformal Thin Sandwich (CTS) Decomposition - {\sl Lagrangian formalism}}
\label{CTS}

Instead of directly treating the extrinsic curvature itself (as in
Section~\ref{EC}), the CTS decomposition \cite{Y99} considers the
evolution of the metric between two neighboring
hypersurfaces\footnote{Advantageously, the method enables an
understanding into the gauge choice and its subsequent evolution
through the kinematic variables, $N$ and $\beta^i$.}. This is
achieved by introducing the time-derivative of the conformal
3-metric, $\tilde{u}_{ij}$,
\begin{equation}
\label{dtuij} \tilde{u}_{ij}  \equiv \partial_{t}
\tilde{\gamma}_{ij},
\end{equation}
$\tilde{\gamma}_{ij}$ defined in (\ref{conformalmetricdefn}), such
that $ \label{uij} u_{ij} \equiv \gamma^{1/3}
\partial_t (\gamma^{-1/3} \gamma_{ij})$\private{\begin{equation} \label{uij} u_{ij}  \equiv \gamma^{1/3}
\partial_t (\gamma^{-1/3} \gamma_{ij}),
\end{equation}} and $\tilde{\gamma}^{ij}\tilde{u}_{ij} \equiv 0$\footnote{This relationship allows for the conformal
metrics on both hypersurfaces to have the same determinant to
first order in $\delta t$}. Using a nontrivial conformal rescaling
of both the lapse\footnote{This scaling is a direct consequence of
the conformal invariance of the lapse antidensity $\alpha$, such
that $\tilde{\alpha} = \alpha$, where $\alpha$ and $\beta^i$ are
undetermined multipliers of the constraints. When the scalar
constraint is satisfied, the ADM action results in the
relationship $\alpha = \tilde{N} \tilde{\gamma}^{-1/2}$, and
therefore, $N \, \equiv \, \Psi^6 \tilde{N}$ is obtained.}, $N \,
\equiv \, \Psi^6 \tilde{N}$, and the tracefree extrinsic
curvature, $A_{ij} \, \equiv \, \Psi^{-2} \tilde{A}_{ij} $, the
kinematic relation (\ref{kinematic1}) simplifies to,
\begin{equation}
\label{AijCTS} A^{ij} =  \Psi^{-10} \tilde{A^{ij}}  \equiv
\frac{\Psi^{-10}}{2 \tilde{N} } \left( ( \tilde{\mathbb{L}}
\beta)^{ij}  -  \tilde{u}^{ij}  \right).
\end{equation}
Equation (\ref{AijCTS}) implies form invariance under conformal
transformations. In summary, the constraint equations assume the
following form in the CTS decomposition,
\begin{eqnarray}
\label{InitialConstraint1CTS}  \widetilde{\nabla}_j \left(
\frac{1}{2 \tilde{N}} (\tilde{\mathbb{L}} \beta)^{ij} \right)
-\widetilde{\nabla}_j \left( \frac{1}{2 \tilde{N}} \tilde{u}^{ij} \right) -
\frac{2}{3} \Psi^6 \widetilde{\nabla}^i K & = & 0 \\
\label{InitialConstraint2CTS} \widetilde{\Delta} \Psi -
\frac{1}{8} \Psi \tilde{R} - \frac{1}{12} \Psi^5 K^2 + \frac{1}{8} \Psi^{-7}
\tilde{A}_{ij} \tilde{A}^{ij} & = & 0
\end{eqnarray}
The set of free data in this case are $(\tilde{\gamma}_{ij},
\tilde{u}^{ij}, K, \tilde{N})$. They solve for the constrained
variables, $\Psi$ and $\beta^i$, in (\ref{InitialConstraint1CTS})
and (\ref{InitialConstraint2CTS}), under appropriate inner and
outer boundary conditions. Thus, the physical initial values,
$\gamma_{ij}$ and $K_{ij}$, follow in a straightforward manner.

We note that in practical computations, the set of four constraint
equations in the CTS decomposition is often complemented by a
fifth coupled elliptical equation relating the conformal lapse,
$\tilde{N}$, to the time derivative of the mean curvature,
$\partial_t K$. By eliminating $R$ from the trace of
(\ref{evolution1}) with (\ref{constraint1}) for a vacuum and after
re-expressing the result in terms of their conformal counterparts,
we obtain,
\begin{equation}
\label{fifthcoupled} \widetilde{\Delta} (\tilde{N} \Psi^7) -
(\tilde{N} \Psi^7) \left[ \frac{1}{8} \tilde{R} + \frac{5}{12} \Psi^4 K^2 +
\frac{7}{8} \Psi^{-8} \tilde{A}_{ij} \tilde{A}^{ij} \right] = -\Psi^5
(\partial_t K - \beta^k \partial_k K),
\end{equation}
or alternatively,
\begin{eqnarray}
\nonumber \widetilde{\Delta} \tilde{N} + 14 \widetilde{\nabla}^i \ln \Psi
\widetilde{\nabla}_i \tilde{N} + \tilde{N} \left[ \frac{3}{4} \tilde{R} +
\frac{1}{6} \Psi^4 K^2 \right. & & \\
\label{fifthcoupled2} \left. - \frac{7}{4} \Psi^{-8}
\tilde{A}_{ij} \tilde{A}^{ij} + 42\widetilde{\nabla}_i \ln \Psi
\widetilde{\nabla}^i \ln \Psi \right] & = & - \Psi^{-2} (\partial_t K -
\beta^k \partial_k K).
\end{eqnarray}
The free data, therefore, consists solely of pairs of variables and
their corresponding velocities, $(\tilde{\gamma}_{ij},
\tilde{u}^{ij}, K,
\partial_t K)$, which is more in line with the Lagrangian approach
of the CTS decomposition than using $(\tilde{\gamma}_{ij},
\tilde{u}^{ij}, K, \tilde{N})$ \cite{Pf04}. An additional
motivation for extending the system of constraint
equations\footnote{Despite its frequent application, uniqueness
and existence proofs do not exist at present for the extended CTS
set of equations. Pfeiffer {\sl et al.} \cite{PfY05} recently
investigated the uniqueness properties of the extended CTS system
though without reaching any firm conclusion. Two distinct
solutions for the same free data based on linearized quadruple
gravitational waves \cite{PfKSS05} were found in the extended
system. For a given physical (conformally scaled) amplitude of the
perturbation, the solution for the physical initial data,
$\gamma_{ij}$ and $K_{ij}$, appears to be unique.} is due to the
natural choice of $\partial_t K = 0$ in practical computations of
quasi-equilibrium binary black initial data
\cite{GGB1,GGB2,Cook02,CookP04}. In
Section~\ref{CTS-Decomposition}, we present the free data for both
the simple and extended CTS formulations, using $\tilde{N}$ and
$\partial_t K$ respectively.

\subsection{Correspondence between the weight function $\sigma$ and lapse function $N$}
\label{weight-and-lapse-functions}

The EC and CTS formalism are equivalent to each other for the
specific choice of $\sigma = 2 N$ and $\tilde{\sigma} = 2
\tilde{N}$. Such an equivalence relationship becomes possible
following the introduction of the weight function, $\sigma$, and by
specifying the particular conformal scaling, $\sigma    \equiv
\Psi^{6}   \tilde{\sigma}$, in the EC decomposition; see
\cite{Pf04,PfY03} for details.

Let us assume a stationary solution to Einstein's equations with
timelike Killing vector $l$ such that $\partial_t g_{ij} = 0$. As
\cite{Pf04,PfY03} details, $\tilde{A}^{ij}_{\mathrm{TT}}$ then
vanishes for stationary spacetimes in the case $\sigma = 2 N$ and
$\tilde{\sigma} = 2 \tilde{N}$ in the EC decomposition, where
$\tilde{A}^{ij}_{\mathrm{TT}}$ is usually identified with the
radiative degrees of freedom. This is generally not the case for
stationary, non-static spacetimes in the previous CTT and PTT
decompositions, which do not include the weight function,
$\sigma$. Note that the standard Bowen-York free data specifies
$\tilde{A}^{ij}_{\mathrm{TT}} = 0$; see \cite{BowY80}.
Alternatively, let us instead examine the particular choice
$\tilde{\sigma} = 1$, corresponding to the CTT
decomposition\footnote{Recall that the CTT decomposition is
frequently used when solving the standard Bowen-York initial data
and is a special case within the overall EC decomposition; see
\cite{BakBrCaLoTa01}.}, in the case $\sigma = 2 N$ and
$\tilde{\sigma} = 2 \tilde{N}$.

In this work, we consider both specific instances; firstly, $\sigma
= 2 N$ and $\tilde{\sigma} = 2 \tilde{N}$ and, secondly,
$\tilde{\sigma} = 1$ (with no assumed relationship between $\sigma$
and $N$). The second case corresponds to the former CTT
decomposition.

\section{Post-Newtonian Free Data}
\label{PN-free-data}

This section presents the proposed free data, namely, $(
\tilde{\gamma}_{ij}, \tilde{A}_{\mathrm{TT}}^{ij}, K,
\tilde{\sigma})$ and $(\tilde{\gamma}_{ij}, \tilde{u}^{ij}, K,
\tilde{N} \, \, \mathrm{or} \, \,
\partial_t K)$, based on post-Newtonian results at 2PN in EC (Hamiltonian)
and CTS (Lagrangian) decompositions respectively. These have been
introduced in Sections~\ref{EC} and \ref{CTS} respectively. We
return in Section~\ref{Proof-of-equivalence} to present our
reasoning behind the choice of the specific form of each of the
components in the above free data.

As discussed in Section~\ref{Conformal-decomposition}, the EC and
CTS methods offer two different perspectives in describing how
each 3-spatial hypersurface is embedded in the full
space-time\footnote{In particular, the CTS method offers insights
into the dynamics of the spacetime.}, but do not differ in their
definitions of the conformal spatial metric,
$\tilde{\gamma}_{ij}$. Hence, the form of proposed conformal
metrics, $\tilde{\gamma}_{ij}$, are, as expected, identical in
both decompositions.

\subsection{EC Decomposition}
\label{EC-Decomposition}

We now present each of the components in the free data $(
\tilde{\gamma}_{ij}, \tilde{A}_{\mathrm{TT}}^{ij}, K,
\tilde{\sigma})$.

\subsubsection{Conformal metric: $\tilde{\gamma}_{ij}$ }
\label{Conformal-metric}

Two possible options for the form of the conformal metric,
$\tilde{\gamma}_{ij}$, are:
\begin{enumerate}
\item[(a)] {\underline{Symmetric-Traceless Form:
$\tilde{\gamma}_{ij}$}}

In this case, $\tilde{\gamma}_{ij}$ assumes the form
\begin{equation}
\label{proposedconformalmetricI1} \tilde{\gamma}_{ij}  =
\delta_{ij}  -  \frac{8 G^2 m_1 m_2}{c^4} \frac{\partial^2 g}{
\partial y_1^{<i} \partial y_2^{j>}}  +  \frac{4 G m_1}{c^4
r_1} v_1^{<i} v_1^{j>}  +  \frac{4 G m_2}{c^4 r_2} v_2^{<i}
v_2^{j>},
\end{equation}
where $m_1$ and $m_2$ refer to each of the two point particle masses
respectively\footnote{The `post-Newtonian' masses, $m_1$ and $m_2$,
are introduced in the post-Newtonian iteration as the coefficients
of Dirac delta functions in the Newtonian density of point-like
particles. They were shown in \cite{B03iv} to agree with the
`geometrostatic' masses associated with the Brill-Lindquist solution
in the time-symmetic instance.}, i.e. the black hole masses in our
model, $\mathbf{y_1}$ and $\mathbf{y_2}$ denote the black hole
positions, $r_1 = |\mathbf{x}-\mathbf{y_1}|$ and $r_2=
|\mathbf{x}-\mathbf{y_2}|$ represent the distances to the black
holes from the field point $\mathbf{x}$, and $r_{12} =
|\mathbf{y_1}-\mathbf{y_{2}}|$ gives the distance between the black
holes. In addition, $\mathbf{v}_1 = d\mathbf{y}_1 /dt$ and
$\mathbf{v}_2 = d\mathbf{y}_2 /dt$ refer to coordinate velocities of
the black holes. The term $\frac{\partial^2 g}{\partial y_1^{<i}
\partial y_2^{j>}}$ comprises all the velocity-independent terms in
(\ref{proposedconformalmetricI1}) and represents the symmetric and
tracefree (STF) projection\footnote{i.e. $Q_{<ij>} \equiv
\frac{1}{2} (Q_{ij} + Q_{ji}) - \frac{1}{3} \delta_{ij} Q_{kk}.$}
of the double derivative of the function $g$ with respect to
$\mathbf{y_1}$ and $\mathbf{y_2}$ respectively. The function $g$
first emerged as an elementary `kernel' for the post-Newtonian
direct iterative works \cite{BDI95,BFP98} and is defined by,
\begin{equation}
\label{g} g (\mathbf{x}; \mathbf{y_1}, \mathbf{y_2})  =  \ln (r_1
+ r_2 + r_{12}).
\end{equation}
It satisfies the Poisson equation in a complete distributional sense
such that,
\begin{equation}
\label{gpoisson} \Delta g  =  \frac{1}{r_1  r_2},
\end{equation}
where $\Delta$ represents the standard flat-space Laplacian with respect to
the field point $\mathbf{x}$. The explicit expression for $\phantom{k}_ig_j$
is, therefore, given by,
\begin{equation}
\label{igj} \phantom{i}_{i}g_{j}  \equiv  \frac{\partial^2 g}{
\partial y_1^{i} \partial y_2^{j}} =  \frac{n_{12}^{i} n_{12}^{j} -
\delta^{ij}}{r_{12} (r_{1} + r_{2} + r_{12})}  +
\frac{(n_{12}^{i}-n_{1}^{i})(n_{12}^{j}+n_{2}^{j})}{(r_{1} + r_{2}
+ r_{12})^{2}},
\end{equation}
where the notations $\mathbf{n_1} = ( \mathbf{x} - \mathbf{y_1}) /
r_1$ and $\mathbf{n_2} = (\mathbf{x} - \mathbf{y_2}) / r_2$ refer
to unit displacement vectors from $\mathbf{x}$ to the black holes,
and $\mathbf{n_{12}} = ( \mathbf{y_1} - \mathbf{y_2}) / r_{12}$ is
the unit displacement from black hole 1 to 2. We henceforth follow
the notation introduced by the post-Newtonian works
\cite{BDI95,BFP98,BIJ02,BFeom}, where $\phantom{i}_{i}g_{j} \equiv
\frac{\partial^2 g}{\partial y_1^{i}
\partial y_2^{j}}$, (\ref{proposedconformalmetricI1}) may be
re-expressed in a fully expanded form as,
\begin{eqnarray} \nonumber \tilde{\gamma}_{ij} & = &
\delta_{ij} - \frac{8 G^2 m_1 m_2}{c^4} \left[
\frac{n_{12}^{<i}n_{12}^{j>}}{r_{12} (r_{1} + r_{2} + r_{12})} +
  \frac{(n_{12}^{<i}-n_{1}^{<i})(n_{12}^{j>}+n_{2}^{j>})}{(r_{1} + r_{2} +
    r_{12})^{2}} \right]\\ \label{proposedconformalmetricI} & & + \frac{4 G
  m_1}{c^4 r_1} v_1^{<i} v_1^{j>} + \frac{4 G m_2}{c^4 r_2} v_2^{<i}
v_2^{j>}.\end{eqnarray} The precise form of
(\ref{proposedconformalmetricI1}) is chosen such that the
post-Newtonian expansion (when $c \rightarrow \infty$) of its
corresponding physical metric $\gamma_{ij}$ is physically
equivalent to the standard post-Newtonian spatial metric in
harmonic coordinates \cite{BFP98} at 2PN order modulo a coordinate
transformation. Such equivalence statements are detailed fully in
Sections~\ref{Statement-Equivalence} and
\ref{Relationship-between-conformal-metrics}. In addition,
(\ref{proposedconformalmetricI1}) is a component of the free data
set $( \tilde{\gamma}_{ij}, \tilde{A}_{\mathrm{TT}}^{ij}, K,
\tilde{\sigma})$. As shown in Section~\ref{Perturbation}, this
free data set refers to a solution, albeit approximately, to the
constraints which differs from the {\sl global} conformally-flat
Bowen-York solution \cite{BowY80} at second post-Newtonian order .

\item[(b)] {\underline{Dirac Coordinates}}

Confining ourselves now to Dirac coordinates, the post--Newtonian derived
conformal metric, $\tilde{\gamma}_{ij}^{\mathit{Dirac}}$, takes the
form\footnote{This may be trivially seen by applying the transverse--traceless
projection tensor, $^{TT}\delta_{ij}^{kl}$, given by \cite{MTW},
\begin{eqnarray}
\nonumber
^{\mathrm{TT}}\delta_{ij}^{kl} & = & \delta_{i}^{k} \delta_{j}^{l} - \delta_{j}^{l} \partial_{ik} \Delta^{-1} - \delta_{i}^{k} \partial_{jl} \Delta^{-1} + \partial_{ik} \Delta^{-1}  (\partial_{jl} \Delta^{-1})\\
\label{TTProjtensor} & & - \frac{1}{2} \delta_{ij} \delta^{kl}  +
\frac{1}{2} \delta_{ij} \partial_{kl} \Delta^{-1} + \frac{1}{2} \delta^{kl}
\partial_{ij} \Delta^{-1} - \frac{1}{2} \partial_{ij} \Delta^{-1}
(\partial_{kl} \Delta^{-1})
\end{eqnarray} on the symmetric-tracefree conformal 3-metric, $\tilde{\gamma}^{ij}$ (\ref{proposedconformalmetricI1}).},
\begin{equation}
\label{proposedconformalmetricdirac2}
\tilde{\gamma}_{ij}^{\sl{Dirac}} = \delta_{ij} - \frac{8 G^2 m_1 m_2}{c^4}
[\phantom{i}_i g_j]^{\mathrm{TT}} + \left[ \frac{4 G m_1}{c^4 r_1} v_1^{i}
v_1^{j} \right]^{\mathrm{TT}} + \left[ \frac{4 G m_2}{c^4 r_2} v_2^{i} v_2^{j}
\right]^{\mathrm{TT}},
\end{equation}
where the term $[ \phantom{i}_i g_j]^{\mathrm{TT}}$ denotes the
transverse--traceless form of the double derivative of the
function $g$ (\ref{g}), with respect to $\mathbf{y_1}$ and
$\mathbf{y_2}$ respectively. The explicit expression of $[
\phantom{i}_i g_j]^{\mathrm{TT}}$ in
(\ref{proposedconformalmetricdirac2}) is
\begin{eqnarray}
\nonumber [ \phantom{i}_i g_j]^{\mathrm{TT}} & = &
\phantom{i}_{(i} g_{j)}   +  \frac{7}{8} \partial_{ij} g  -
\frac{3}{16} \partial_{ij} \left( \frac{r_1+r_2}{r_{12}} \right)
\\ \nonumber & &  +  \frac{1}{4} \partial_{(i}
\partial_{2j)} \left( \frac{r_1 - r_2}{r_{12}} \right)  +
\frac{1}{4} \partial_{(j} \partial_{1i)} \left( \frac{r_2 -
r_1}{r_{12}} \right)  +  \frac{1}{96} \partial_{ij} D \left(
\frac{r_1^3 + r_2^3}{ r_{12}} \right) \\ \label{gTT} & &  -
\frac{1}{8} \delta^{ij} \left( 2 D g - D (\frac{r_1+r_2}{r_{12}})
\right),
\end{eqnarray}
where we refer to the notation used in \cite{BFP98,B03iv} for $D
\equiv \frac{\partial^2}{\partial y_1^i
\partial y_2^i}$, and the expressions subsequently derivable,
\begin{eqnarray}
\label{Dg}
D g & = & \frac{1}{2 r_1 r_2}  -   \frac{1}{2 r_1 r_{12}}  -   \frac{1}{2 r_2 r_{12}}, \\
\label{dijigj}
\partial_{ij} \left( \frac{\partial^2 g}{\partial y_1^i \partial y_2^j}
\right) & = & D \left( \frac{1}{2 r_1 r_2} + \frac{1}{2 r_1 r_{12}} +
\frac{1}{2 r_2 r_{12}} \right).
\end{eqnarray}
Similarly, the transverse-traceless velocity-dependent terms,
$\left[ \frac{4 G m_1}{c^4 r_1} v_1^{i} v_1^{j}
\right]^{\mathrm{TT}}$ (and $1 \leftrightarrow 2$), in
(\ref{proposedconformalmetricdirac2}) are given by
\begin{eqnarray}
\nonumber \left[ \frac{4 G m_1}{c^4 r_1} v_1^{<i} v_1^{j>}
\right]^{\mathrm{TT}} & = & \frac{G m_1}{c^4 r_1} \left( \frac{1}{2} v_1^{(i}
v_1^{j)} + \delta^{ij} ( - \frac{5}{4} (n_1 v_1)^2 + \frac{1}{4}
\mathbf{v_1}^2 ) + 3 (n_1 v_1) n_1^{(i} v_1^{j)} \right. \\
\label{hdashTT} & & \left.  +  n_1^{ij} \left( \frac{3}{4} (n_1
v_1)^2 - \frac{5}{4} \mathbf{v_1}^2 \right) \right)
\end{eqnarray}
and $1 \leftrightarrow 2$, where $1 \leftrightarrow 2$ denotes the
exchange of particle labels 1 and 2. In more detail, we have
chosen in (\ref{proposedconformalmetricdirac2}) to fix the spatial
coordinates, $x^i$, of the conformal 3-metric
$\tilde{\gamma}_{ij}^{\mathit{Dirac}}$ on each hypersurface,
$\Sigma_t$, in the generalized Dirac gauge, as introduced in
\cite{BGGN04}. The covariant generalized Dirac gauge is defined
there as,
\begin{equation}
\label{DiracGaugeGen} {\cal D}_{j} \left[ \left( \frac{\gamma}{f}
\right)^{1/3} \ \gamma^{ij \, \mathit{Dirac}} \right] = 0,
\end{equation}
where $f$ and ${\cal D}$ denote, respectively the determinant and
the unique covariant derivative with respect to a flat metric,
$f_{ij}$ (in an arbitrary coordinate system). Equivalently,
\eqref{DiracGaugeGen} may be expressed in terms of the conformal
metric as $\frac{\partial}{\partial x^{j}} \tilde{\gamma}^{ij \,
\mathit{Dirac}} = 0$, where the flat metric $f^{ij} \equiv
\delta^{ij}$ assumes Minkowski coordinates. \private{
\begin{equation}
\label{DiracGaugeGenConf} \frac{\partial}{\partial x^{j}}
\tilde{\gamma}^{ij \, \mathit{Dirac}} = 0,
\end{equation}}

Finally, from (\ref{gTT}) and (\ref{hdashTT}), the explicit form of
(\ref{proposedconformalmetricdirac2}) is given as,
\begin{eqnarray}
\nonumber \tilde{\gamma}_{ij}^{\sl{Dirac}} - \delta_{ij}& = &\frac{G^2 m_1
m_2}{c^4} \left[ \delta^{ij} \left( - \frac{5 r_1}{ 8 r_{12}^3} - \frac{15}{8
r_1 r_{12}} + \frac{5 r_1^2}{8 r_{12}^3 r_2} + \frac{1}{(r_1 +r_2 +r_{12})^2}
\left(1 \right. \right. \right. \\ \nonumber & & \left. \left. \left. +
\frac{r_1}{r_{12}}+\frac{r_{12}}{r_{1}} -\frac{r_{1}}{r_{2}}-\frac{r_{1}^2}{
r_{2} r_{12}}+\frac{r_{12}^2}{2 r_{1} r_{2}} \right) + \frac{1}{(r_1 +r_2
+r_{12})} ( - \frac{7}{r_{1}} + \frac{2}{r_{12}} )\right) \right. \\ \nonumber
& & \left. + n_1^{i} n_1^j \left( \frac{r_1}{8 r_{12}^3}+\frac{11}{8 r_1
r_{12}} - \frac{r_2^2}{8 r_1 r_{12}^3} + \frac{7}{(r_1 +r_2 +r_{12})^2}
+\frac{7}{r_1(r_1 +r_2 +r_{12})} \right) \right. \\ \nonumber & & + n_1^{(i}
n_{12}^{j)} \left( - \frac{7}{2 r_{12}^2} + \frac{8}{(r_1 +r_2 +r_{12})^2}
\right) + n_{12}^{i} n_{12}^{j} \left( - \frac{4}{(r_1+r_2 +r_{12})^2} \right.
\\ \nonumber & & \left. \left. -\frac{4}{r_{12}(r_1 +r_2 +r_{12})} \right) +
\frac{11 n_1^{(i}n_2^{j)}}{(r_1 +r_2 +r_{12})^2} \right] \\
\nonumber & & + \frac{G m_1}{c^4 r_1} \left[ \frac{v_1^{(i}
v_1^{j)}}{2} + \delta^{ij} \left( - \frac{5 (n_1 v_1)^2 }{4} +
\frac{\mathbf{v_1}^2}{4} \right) + 3 (n_1 v_1) n_1^{(i} v_1^{j)}
\right. \\ \label{gammatildeTT} & & \left. + n_1^{i} n_1^j \left(
\frac{3(n_1 v_1)^2}{4} - \frac{5\mathbf{v_1}^2}{4} \right) \right]+
1 \leftrightarrow 2.
\end{eqnarray}

\end{enumerate}
\subsubsection{Extrinsic Curvature: $\tilde{A}^{ij}_{\mathrm{TT}}$ and $K$}

The conformal symmetric transverse-tracefree tensor component of the
extrinsic curvature, $\tilde{A}^{ij}_{\mathrm{TT}}$, is given by,
\begin{equation}
\label{AtildeijTT1} \tilde{A}^{ij}_{\mathrm{TT}} = 0,
\end{equation}
for both possible values of $K$: maximal hypersurface $K = 0$ and
the post-Newtonian mean curvature $K^{\mathrm{2PN}}$ derived at
2PN using results in \cite{BFP98} and given by,
\begin{equation}
\label{2PNK} K^{\mathrm{2PN}} = \frac{G m_1 (n_1 v_1)}{c^3 r_1^2} +
\frac{G m_2 (n_2 v_2)}{c^3 r_2^2} + {\cal O}
\left(\frac{1}{c^5}\right).
\end{equation}
The choice of $\tilde{A}^{ij}_{\mathrm{TT}}=0 $ and $K=0$ are in
agreement at 2PN order with their counterparts in the standard
Bowen-York solution, given in \cite{BowY80}. Maximal slicing is
considered advantageous primarily due to its `singularity
avoiding' feature during evolution; the slicing causes the lapse,
$N$, to collapse to zero in the region where a physical
singularity exists; see \cite{Al04b} for a general discussion.

\subsubsection{Weight function: $\tilde{\sigma}$}

As is mentioned in Section~\ref{weight-and-lapse-functions}, we
consider two specific instances; a) $\sigma = 2 N$ and
$\tilde{\sigma} = 2 \tilde{N}$ and, b) $\tilde{\sigma} = 1$.

\begin{enumerate}

\item[(a)] \underline{$\sigma = 2N$ and $\tilde{\sigma} = 2
\tilde{N}$}

This concerns the class of solutions where the EC and CTS
decompositions are seen to be equivalent. Depending on the choice
in mean curvature, either $K=0$ or $K = K^{\mathrm{2PN}}$, and
whether the conformal 3-metric, $\tilde{\gamma}_{ij}$, is
specified in symmetric trace-free form or in Dirac coordinates
(following the relationship $N = \Psi^6 \tilde{N}$), the conformal
weight function based on 2PN results,
$\tilde{\sigma}^{\mathrm{2PN}}$, can take four different forms.
These are:

\begin{enumerate}

\item[i)] For a maximal hypersurface $K=0$,
$\tilde{\sigma}|_{K=0}$ is given as,
\begin{eqnarray} \nonumber \tilde{\sigma}^{\mathrm{STT}}|_{K=0} -2 \equiv 2
  \tilde{N}^{\mathrm{STT}}|_{K=0} -2 & =&- \frac{8 G m_1}{c^2 r_1} + \frac{1}{c^4} \left[ \frac{G m_1}{r_1} [
      3 (n_1 v_1)^2 - 7 v_1^2 ] + \frac{35 G^2 m_1^2}{2 c^4 r_1^2} \right.
      \\ \label{sigmaECSTTmaxhyp} & & \left.   +
      G^2 m_1 m_2 \left( \frac{18}{r_1 r_2} + \frac{10}{r_{12} r_2} \right)
      \right] + 1 \leftrightarrow 2,
\end{eqnarray}
in a symmetric trace-free coordinate system.

\item[ii)] Similarly, in Dirac coordinates,
\begin{eqnarray}
\nonumber \tilde{\sigma}^{\sl{Dirac}} |_{K=0} -2  \equiv 2
  \tilde{N}^{\sl{Dirac}}|_{K=0} -2 & = & - \frac{8 G
  m_1}{c^2 r_1} + \frac{1}{c^4} \left[ - \frac{6 G m_1}{r_1}  v_1^2 + \frac{35
  G^2 m_1^2}{2 c^4 r_1^2} \right. \\ \nonumber  & & \left.+
      G^2 m_1 m_2 \left( \frac{35}{ 2 r_1 r_2} + \frac{5}{ r_{12} r_2}
      \right)
      \right] + 1 \leftrightarrow 2. \\ \label{sigmaECDiracmaxhyp}
\end{eqnarray}

\item[iii)] For the 2PN derived mean curvature
$K^{\mathrm{2PN}}$, $\tilde{\sigma}|_{K^{\mathrm{2PN}}}$ is given
as,
\begin{eqnarray} \nonumber
  \tilde{\sigma}^{\mathrm{STT}}|_{K^{\mathrm{2PN}}} -2  \equiv 2
  \tilde{N}^{\mathrm{STT}}|_{K^{\mathrm{2PN}}} -2  & =& - \frac{8 G m_1}{c^2
      r_1} + \frac{1}{c^4} \left[ \frac{G m_1}{r_1} [4 (n_1 v_1)^2 - 8 v_1^2 ]
      + \frac{35 G^2 m_1^2}{2 c^4 r_1^2} \right. \\ \nonumber & &
      \left.+ G^2 m_1 m_2 \left( \frac{18}{r_1 r_2} + \frac{21}{2 r_{12} r_2} +
      \frac{r_1}{2 r_{12}^3} - \frac{r_1^2}{2 r_{12}^3 r_2} \right) \right] \\
      \nonumber & & \label{sigmaECSTT2PN} +
      1 \leftrightarrow 2,
\end{eqnarray}
in the symmetric trace-free coordinate system.

\item[iv)] Similarly, in Dirac coordinates,
\begin{eqnarray} \nonumber
\tilde{\sigma}^{\sl{Dirac}}|_{K^{\mathrm{2PN}}} -2  \equiv 2
\tilde{N}^{\sl{Dirac}}|_{K^{\mathrm{2PN}}} -2  & = & - \frac{8 G
m_1}{c^2 r_1} + \frac{1}{c^4} \left[ \frac{G m_1}{r_1} [ (n_1
      v_1)^2 - 7 v_1^2]  + \frac{35 G^2 m_1^2}{2 c^4 r_1^2}
      \right. \\ \nonumber & & \left. +
      G^2 m_1 m_2 \left( \frac{39}{2 r_1 r_2} + \frac{11}{2 r_{12} r_2} +
      \frac{r_1}{2 r_{12}^3} - \frac{r_1^2}{2 r_{12}^3 r_2} \right) \right] \\ \nonumber
 & & \label{sigmaECDirac2PN} + 1 \leftrightarrow 2.
\end{eqnarray}

\end{enumerate}

\item[(b)] {\underline{$\tilde{\sigma}^{\mathrm{CTT}}$ in the CTT
decomposition}}

In this case, the conformal weight function
$\tilde{\sigma}^{\mathrm{CTT}} = 1$.

\end{enumerate}

\subsection{CTS Decomposition}
\label{CTS-Decomposition}

We now present each of the components in the free data
$(\tilde{\gamma}_{ij}, \tilde{u}^{ij}, K, \tilde{N} \, \,
\mathrm{or} \, \,
\partial_t K)$.

\subsubsection{Conformal metric: $\tilde{\gamma}_{ij}$}

Due to the identical nature of the conformal decompositions of the
3-metric $\gamma_{ij}$, the post-Newtonian motivated conformal
metric in symmetric-tracefree form, $\tilde{\gamma}_{ij}$, and in
Dirac coordinates, $\tilde{\gamma}_{ij}^{{\mathit Dirac}}$, are
given by (\ref{proposedconformalmetricI1}) and
(\ref{proposedconformalmetricdirac2}) respectively.

\subsubsection{Time derivative of the conformal metric, $\tilde{u}_{ij}$, and mean curvature $K$}

Following definition (\ref{dtuij}) and using either $\gamma_{ij}$
(\ref{proposedconformalmetricI1}) or $\tilde{\gamma}_{ij}$
(\ref{proposedconformalmetricdirac2}), we propose that the
post-Newtonian time derivative of the conformal metric,
$\tilde{u}_{ij}$, adopts the form,
\begin{equation}
\label{uij} \tilde{u}^{ij} = 0,
\end{equation}
at 2PN for both a maximal hypersurface, $K=0$, and the 2PN derived
mean curvature $K^{\mathrm{2PN}}$, given by (\ref{2PNK}). The
particular choice in $\tilde{u}_{ij} = 0$ and $K = 0$ are in
agreement with quasi-stationary initial conditions\footnote{Note
that helical symmetry is exact with respect to Newtonian and 2PN
gravity.} \cite{GGB1,GGB2,Cook02,CookP04}.

\subsubsection{Conformal Lapse $\tilde{N}$, or time derivative of mean curvature, $\partial_t K$}

As discussed in Section~\ref{CTS}, we give both the conformal lapse,
$\tilde{N}$, and the time derivative of mean curvature, $\partial_t
K$, depending on whether the simple, or the extended, CTS system of
constraint equations is to be used.

\begin{enumerate}

\item [(a)] {\underline{Conformal Lapse $\tilde{N}$}}

Thanks to the choice in both mean curvature, $K=0$ or
$K^{\mathrm{2PN}}$, and the preferred spatial coordinate system of
$\tilde{\gamma}_{ij}$ or $\tilde{\gamma}_{ij}^{\sl{Dirac}}$, there
are four different possibilities for the 2PN based conformal lapse,
$\tilde{N}$. These are:

\begin{enumerate}

\item[i)] For a maximal hypersurface $K=0$, $\tilde{N}|_{K=0}$ is given as,
\begin{eqnarray} \nonumber \tilde{N}^{\mathrm{STT}} |_{K=0}
  - 1 & =&- \frac{4 G m_1}{c^2 r_1} + \frac{1}{c^4} \left[ \frac{G m_1}{r_1} \left(
      \frac{3}{2} (n_1 v_1)^2 - \frac{7 v_1^2}{2} \right) + \frac{35 G^2 m_1^2}{4 c^4 r_1^2} \right.
      \\ \label{NSTTmaxhyp} & & \left.   +
      G^2 m_1 m_2 \left( \frac{9}{r_1 r_2} + \frac{5}{r_{12} r_2} \right)
      \right]+ 1 \leftrightarrow 2,
\end{eqnarray}
in the symmetric trace-free coordinate system.

\item[ii)] Similarly, in Dirac coordinates,
\begin{eqnarray}
\nonumber \tilde{N}^{\sl{Dirac}} |_{K=0} - 1 & = & - \frac{4 G
  m_1}{c^2 r_1} + \frac{1}{c^4} \left[ - \frac{3 G m_1}{r_1}  v_1^2 + \frac{35
  G^2 m_1^2}{4 c^4 r_1^2} \right. \\ \label{NDiracmaxhyp} & & \left.+
      G^2 m_1 m_2 \left( \frac{35}{ 4 r_1 r_2} + \frac{5}{2 r_{12} r_2} \right)
      \right] + 1 \leftrightarrow 2.
\end{eqnarray}

\item[iii)] For the 2PN derived mean curvature $K^{\mathrm{2PN}}$,
$\tilde{N}|_{K^{\mathrm{2PN}}}$ is given as,
\begin{eqnarray} \nonumber
  \tilde{N}^{\mathrm{STT}}|_{K^{\mathrm{2PN}}} -1 & =&- \frac{4 G m_1}{c^2
      r_1} + \frac{1}{c^4} \left[ \frac{G m_1}{r_1} [2 (n_1 v_1)^2 - 4 v_1^2 ]
      + \frac{35 G^2 m_1^2}{4 c^4 r_1^2} \right. \\ \nonumber & &
      \left.+ G^2 m_1 m_2 \left( \frac{9}{r_1 r_2} + \frac{21}{4 r_{12} r_2} +
      \frac{r_1}{4 r_{12}^3} - \frac{r_1^2}{4 r_{12}^3 r_2} \right) \right] \\
      \label{NSTT2PN} & & +
      1 \leftrightarrow 2,
\end{eqnarray}
in the symmetric trace-free coordinate system.

\item[iv)] Similarly, in Dirac coordinates,
\begin{eqnarray}
\nonumber \tilde{N}^{\sl{Dirac}}|_{K^{\mathrm{2PN}}} -1 & = & -
\frac{4 G m_1}{c^2 r_1} + \frac{1}{c^4} \left[ \frac{G m_1}{r_1} \left( \frac{ (n_1
      v_1)^2}{2} - \frac{7 v_1^2}{2} \right) + \frac{35 G^2 m_1^2}{4 c^4 r_1^2}
      \right. \\ \nonumber & & \left. +
      G^2 m_1 m_2 \left( \frac{39}{4 r_1 r_2} + \frac{11}{4 r_{12} r_2} +
      \frac{r_1}{4 r_{12}^3} - \frac{r_1^2}{4 r_{12}^3 r_2} \right) \right] \\
\label{NDirac2PN} & & + 1 \leftrightarrow 2.
\end{eqnarray}

\end{enumerate}

\item [(b)] {\underline{Time derivative of mean curvature, $\partial_t K$}}

Since there exist two alternative slicing possibilities for our PN
free data, $K=0$ or $K \equiv K^{\mathrm{2PN}}$, their time
derivatives are given respectively by: i) $\partial_{t} K = 0$, or
ii)
\begin{eqnarray} \nonumber
\partial_t K^{\mathrm{2PN}} & = & \frac{G m_1}{c^4 r_1^3} \left( 3
(n_1 v_1)^2 - \mathbf{v_1}^2 \right) + \frac{G^{2} m_1 m_2}{c^4
r_{12}} \left( \frac{1}{2 r_1 r_{12}^2} + \frac{1}{2 r_1^3} - \frac{r_1^2} { 2
r_{12}^2 r_2^3} \right) \\ \label{dtK2PN} & & + 1 \leftrightarrow 2 + {\cal O} \left( \frac{1}{c^5} \right).
\end{eqnarray}

\end{enumerate}

\section{Proof of equivalence with Post-Newtonian results}
\label{Proof-of-equivalence}

This section presents our reasoning behind the choice of free data
discussed in Section~\ref{PN-free-data} and, in particular, their
full agreement with post-Newtonian results at 2PN order. Note that
the free data are also chosen such that they satisfy the
constraints, albeit approximately. In illustrating this agreement,
we state first in Section~\ref{Statement-Equivalence} our central
result exhibiting the ``near zone" behavior of the proposed
solution and outline then its derivation in
Section~\ref{Relationship-between-conformal-metrics} by
considering the form of the proposed PN-derived conformal
3-metrics, $\tilde{\gamma}_{ij}$ and
$\tilde{\gamma}_{ij}^{\sl{Dirac}}$. In addition, motivations for
the specific form of our PN free data are given in
Section~\ref{Perturbation} by considering the lowest order
perturbation of the Bowen-York solution. This allows us to
investigate both the global structure and near-zone solution of
our proposed PN based solution. Finally, in
Section~\ref{Extrinsic-Curvature-and-Maximal-Slicing}, we show
that our results imply that the physical extrinsic curvature from
the standard Bowen-York solution, $K_{ij}^{\mathrm{B-Y}}$, and
post-Newtonian derived counterpart, $K_{ij}^{\mathrm{2PN}}$ are
physically equivalent.

\subsection{Statement of Equivalence}
\label{Statement-Equivalence}

The statement of equivalence presented below is a generalization of
Theorem 1 in \cite{B03iv} for stationary black holes to moving black
holes with $\mathbf{v_1}, \mathbf{v_2} \neq 0$.

The conformal metrics, $\tilde{\gamma}_{ij}$ and
$\tilde{\gamma}_{ij}^{\mathit{Dirac}}$, are chosen in such a way
that a post-Newtonian expansion (when $c \rightarrow \infty$) of
their corresponding physical metrics, $\gamma_{ij} = \Psi^4
\tilde{\gamma}_{ij}$ and $\gamma_{ij}^{\mathit{Dirac}}  =  ( \,
\Psi^{\mathit{Dirac}} \, )^4 \,
\tilde{\gamma}_{ij}^{\mathit{Dirac}}$, are physically equivalent to
the standard post-Newtonian spatial metric in harmonic coordinates
at 2PN order, that is, they differ only by a change of coordinates.
This can be stated as,
\begin{equation} \label{pNandconfmetricI} \gamma_{ij} =
g^{\mathrm{2PN}}_{ij} +  \partial_i \xi_j +  \partial_j \xi_i  +
\mathcal{O} \left(\frac{1}{c^5}\right),
\end{equation}
A similar statement applies to Dirac coordinates. In other words,
\begin{equation} \label{pNandconfmetricdirac}
\gamma_{ij}^{\mathit{Dirac}} = g^{\mathrm{2PN}}_{ij} +  \partial_i
\xi_j^{\mathit{Dirac}} +
\partial_j \xi_i^{\mathit{Dirac}} +  \mathcal{O}
\left(\frac{1}{c^5}\right).
\end{equation}
In (\ref{pNandconfmetricI}) and (\ref{pNandconfmetricdirac}),
$g^{\mathrm{2PN}}_{ij}$ represents the spatial metric in harmonic
coordinates truncated at 2PN order \cite{BFP98}. The remainder
$\mathcal{O}\left(\frac{1}{c^5}\right)$ accounts for neglected 2.5PN
and higher-order terms. The change in coordinates is specified by
the unique spatial gauge transformation, $\xi^i$ or
${\xi^i}^{\sl{Dirac}}$, depending on the preferred coordinate
system.

Additionally, our PN data obeys the following relationships for
either maximal slicing, $K=0$,
\begin{eqnarray} \label{pNgamma0i} \gamma_{0i} & = &
g^{\mathrm{2PN}}_{0i} +  \partial_0 \xi_i +  \partial_i \xi_0 +
\mathcal{O} \left(\frac{1}{c^5}\right), \\ \label{pNgamma00} \gamma_{00} & = & g^{\mathrm{2PN}}_{00} +  2
\partial_0 \xi_0 +  \mathcal{O} \left(\frac{1}{c^5}\right),
\end{eqnarray}
\noindent or, mean curvature $K_{2PN}$,
\begin{eqnarray}  \label{pNgamma0i'} \gamma'_{0i} & = &
g^{\mathrm{2PN}}_{0i} + \partial_0 \xi'_i + \partial_i \xi'_0 +
\mathcal{O} \left(\frac{1}{c^5}\right), \\ \label{pNgamma00'}
\gamma'_{00} &=& g^{\mathrm{2PN}}_{00} + 2 \partial_0 \xi'_0 +
\mathcal{O} \left(\frac{1}{c^5}\right),
\end{eqnarray}
where $\xi'_i = \xi_i + \mathcal{O} (1/c^5)$ and
$g^{\mathrm{2PN}}_{00}$ and $g^{\mathrm{2PN}}_{0i}$ represent the
$00$th and $0i$th component of the full spacetime metric in
harmonic coordinates \cite{BFP98}.

The vector, $\xi^{\mu}$, which is determined by
(\ref{pNandconfmetricI})-(\ref{pNgamma00'}), represents {\sl unique
infinitesimal gauge transformation} (i.e. $x^{\mu} \rightarrow
x'^{\mu} = x^{\mu} + \xi^{\mu} (x^{\nu}) $, where $\{x^{\mu}\} $ and
$\{x'^{\mu}\} $ are two general coordinate systems) at 2PN order and
is explicitly given by the following components,
\begin{eqnarray}
\label{xi01} \xi_{0} & = &  \frac{G m_{1}}{2 c^{3}}  (
n_{1} v_{1} ) +  \frac{G m_{2}}{2 c^{3}} ( n_{2}
v_{2} ) + {\cal O} \left( \frac{1}{c^4} \right), \\
\label{xi01'} \xi'_{0} & = &  0 + {\cal O} \left( \frac{1}{c^4} \right),
\end{eqnarray}
for either mean curvature $K=0$ or $K_{2PN}$ respectively,
\begin{eqnarray} \label{xiiSTT}
\xi_i & = &  \frac{G^2 m_1^2}{4 c^4} \partial_i \ln r_1  +
\frac{G^2 m_2^2}{4 c^4} \partial_i \ln r_2  +  \mathcal{O}
\left(\frac{1}{c^5}\right),
\end{eqnarray}
for the symmetric trace-free coordinate system and, similarly, in
Dirac coordinates,
\begin{eqnarray} \nonumber
\xi_i^{\mathit Dirac} & = & \frac{G^2 m_1^2}{4 c^4} \partial_i \ln
r_1 - \frac{7 G^2 m_1 m_2}{2 c^4} \frac{n_1^i}{(r_1+r_2+r_{12})} +
\frac{3 G^2 m_1 m_2}{8 c^4}
\partial_{i} \left( \frac{r_1 + r_2}{r_{12}} \right) \\ \nonumber &
& - \frac{G^2 m_1 m_2}{c^4}
\partial_{1i} \left( \frac{r_2 - r_1}{r_{12}} \right)   -  \frac{G^2 m_1 m_2}{48 c^4} \partial_{i} \mathrm{D} \left( \frac{r_1^3+r_2^3}{r_{12}} \right) -  \frac{2 G
m_1}{c^4} \partial_k \left( v_1^{(i} v_1^{k)} r_1 \right)\\
\label{xiiDirac} & &   + \frac{G m_1}{24 c^4}
\partial_{ikl} \left( v_1^{(k} v_1^{l)} r_1^3 \right) + \frac{G m_1}{2 c^4}
\partial_i (\mathbf{v_1}^2 r_1) + 1 \leftrightarrow 2 +
\mathcal{O}\left(\frac{1}{c^5}\right).
\end{eqnarray}

\subsection{Relationship between the conformal metrics, $\tilde{\gamma}_{ij}$ and $\tilde{\gamma}_{ij}^{\mathit{Dirac}}$, and the post--Newtonian
metric}
\label{Relationship-between-conformal-metrics}

In order to demonstrate results (\ref{pNandconfmetricI}) and
(\ref{pNandconfmetricdirac}), let us consider initially a
post-Newtonian iteration of the Hamiltonian constraint,
(\ref{InitialConstraint2EC}) and (\ref{InitialConstraint2CTS}),
for proposed conformal metric in symmetric-trace-free form,
$\tilde{\gamma}_{ij}$, at 2PN order,
\begin{equation} \label{DeltaPsi}
\Delta \Psi =  - \frac{G^2 m_1 m_2}{c^4} \partial_{ij}
\left(\phantom{i}_{<i}g_{j>} \right) +  \frac{G m_1}{2 c^4 r_1^3}
(n_1 v_1)  +  \frac{ G m_2}{2 c^4 r_2^3}  (n_2 v_2)  + {\cal O}
\left( \frac{1}{c^5} \right),
\end{equation}
where we note the absence of any contributing terms from the
extrinsic curvature, $K_{ij}$. As
Section~\ref{Extrinsic-Curvature-and-Maximal-Slicing} shows, this
is unsurprising considering that the lowest order term in the
extrinsic curvature, $K_{ij}$, appears at ${\cal O}
(\frac{1}{c^3})$. Therefore, the quadratic terms of the conformal
symmetric transverse-trace-free tensor
$\tilde{A}^{ij}_{\mathrm{TT}}$, and mean curvature, $K$, occur at
order ${\cal O} (\frac{1}{c^6})$ in (\ref{InitialConstraint2EC})
and (\ref{InitialConstraint2CTS}). The most general solution to
(\ref{DeltaPsi}), in the full distributional sense, for the
conformal factor, $\Psi$, is given by,
\begin{eqnarray}
\nonumber \Psi & = & \psi  -  \frac{G^2 m_1 m_2}{2 c^4} \mathrm{D}
\left(  \frac{g}{3} + \frac{r_1+r_2}{2 r_{12}} \right)  - \frac{G
m_1}{12 c^4 r_1} \left( 3 (n_1 v_1)^2 - \mathbf{v_1}^2 \right) \\
\label{Psi2pn} & &  -  \frac{G m_2}{12 c^4 r_2} \left( 3 (n_2 v_2)^2
- \mathbf{v_2}^2 \right) +  {\cal O} \left( \frac{1}{c^5} \right),
\end{eqnarray} $\psi$ being the solution of the homogenous equation, $\Delta \psi =
0$. This result, when instantiated with $\mathbf{v_1} = \mathbf{v_2}
= 0$, is consistent with the time-symmetric result of \cite{B03iv},
given there by Eqns.~(2.6) and (2.7). The explicit form of $\psi$ is
then obtained by comparing (\ref{Psi2pn}) with the post-Newtonian
spatial metric, $g_{ij}^{\mathrm{2PN}}$ (i.e. Eqn.~(7.2) in
\cite{BFP98}) in the form,
\begin{eqnarray}
\label{psi} \psi & = & 1  +  \frac{G m_1}{2 c^2 r_1} \left(  1  -
\frac{G m_2}{2 c^2 r_{12}} +  \frac{\mathbf{v_1}^2}{2 c^2} \right)
+ \frac{G m_2}{2 c^2 r_2}  \left(  1 -  \frac{G m_1}{2 c^2 r_{12}}
+ \frac{\mathbf{v_2}^2}{2 c^2} \right) +  {\cal O} \left(
\frac{1}{c^5} \right).
\end{eqnarray}
Since further `homogeneous' terms $\sim 1/r_1$ and $\sim 1/r_2$ can
be added to $\Psi$, without affecting (\ref{DeltaPsi}), for
consistency, we specify that $\Psi$ must satisfy (\ref{DeltaPsi}) in
a strict distributional sense; see Section III A in \cite{B03iv}.
Therefore, we do not allow the addition of such terms, here and
henceforth, to our solutions for Poisson-type equations.

Interestingly, by re-expressing $\psi$, (\ref{psi}), in terms of
`Brill-Lindquist-like' constants, $\alpha_1$ and $\alpha_2$,
\begin{eqnarray} \label{psiBLlike} \psi & \equiv & 1 +
\frac{\alpha_1}{r_1}+\frac{\alpha_2}{r_2},
\end{eqnarray}
where $\alpha_1$ and $\alpha_2$ are determined to the relative 2PN
accuracy from (\ref{psi}) as,
\begin{eqnarray}
\label{alpha1} \alpha_1 & = & \frac{G m_1}{2 c^2} \left( 1  -
\frac{G m_2}{2 c^2 r_{12}} +  \frac{\mathbf{v_1}^2}{2 c^2}  +
\mathcal{O} \left( \frac{1}{c^3} \right) \right)   \, \, \, \,  \mathrm{and} \,
\, \, \, 1 \leftrightarrow  2,
\end{eqnarray}
we recover the original Brill-Lindquist conformal factor,
$\psi_{\mathrm{BL}}$, for the time-symmetric instance
$\mathbf{v_1} = \mathbf{v_2} = 0$ at 2PN order\footnote{$\psi_{\mathrm{BL}}$ is given here for completeness as,
\begin{eqnarray} \nonumber \psi^{\mathrm{B-L}} & = & 1 + \frac{\alpha_1^{\mathrm{B-L}}}{r_1}
 + \frac{\alpha_2^{\mathrm{B-L}}}{r_2} \\ \label{psiBL} & = &
1 + \frac{G m_1}{2 c^2 r_1} \left( 1 - \frac{G m_2}{2 c^2 r_{12}}
\right) + \frac{G m_2}{2 c^2 r_2} \left( 1 - \frac{G m_1}{2 c^2
r_{12}} \right) + \mathcal{O} \left( \frac{1}{c^6} \right).
\end{eqnarray}}.

In contrast, when following a similar procedure for Dirac
coordinates, we find that the conformal factor $\Psi^{\sl{Dirac}}$
is equivalent to the term $\psi$, (\ref{psi}), i.e.
\begin{eqnarray} \nonumber
\Psi^{\sl{Dirac}} \equiv \psi & = & 1 + \frac{\alpha_1}{r_1} +
\frac{\alpha_2}{r_2} \\ \nonumber & = & 1 + \frac{G m_1}{2 c^2
r_1} \left( 1 - \frac{G m_2}{2 c^2 r_{12}} +
\frac{\mathbf{v_1}^2}{2 c^2} \right) + \frac{G m_2}{2 c^2 r_2}
\left( 1 - \frac{G m_1}{2 c^2 r_{12}} + \frac{\mathbf{v_2}^2}{2
c^2} \right) \\ \label{psiDIRAC} & & + \mathcal{O} \left(
\frac{1}{c^5} \right).
\end{eqnarray}
Therefore, the deviation of our post-Newtonian motivated solution
from the conformally flat solution in the Dirac gauge manifests
itself solely as a perturbation in the conformal metric,
$\tilde{\gamma}_{ij}^{\sl{Dirac}}$ (\ref{proposedconformalmetricdirac2}), and not to the
`Brill-Lindquist'-like conformal factor, $\psi$ (\ref{psi}), at the 2PN
order.

\subsection{Perturbation of a general global conformally--flat
solution}
\label{Perturbation}

This section considers the lowest order perturbation of the standard
conformally-flat Bowen-York solution. It gives an insight into the
specific form of the conformal 3 metric, $\tilde{\gamma}_{ij} $.
Note that we only consider the symmetric trace-free form of the
proposed conformal metric, $\tilde{\gamma}_{ij}$
(\ref{proposedconformalmetricI1}). This is because, as discussed in
Section~\ref{Relationship-between-conformal-metrics}, the PN based
conformal metric in Dirac coordinates,
$\tilde{\gamma}_{ij}^{\sl{Dirac}}$
(\ref{proposedconformalmetricdirac2}), incorporates fully the
characteristics of the 2PN metric itself and does not incur a
perturbative change to the Brill-Lindquist-like conformal factor,
$\psi$ (\ref{psi}).

Let us introduce formally a metric perturbation, $h_{ij} $, to the
conformal metric $\tilde{\gamma}_{ij} $, where we interpret the
proposed 2PN motivated conformal metric
(\ref{proposedconformalmetricI1}) as a perturbation to the
Bowen-York solution, i.e.,
\begin{equation}
\label{perturbationconformalflatness} \widetilde{\gamma}_{ij} =
\delta_{ij}   +   h_{ij},
\end{equation}
where $h = h_{ii} = 0 $ and $h_{ij}$ is explicitly given from
(\ref{proposedconformalmetricI1}) as,
\begin{equation}
\label{perturbationconformalflatness2} h_{ij} = -  \frac{8 G^2 m_1
m_2}{c^4} \frac{\partial^2 g}{ \partial y_1^{<i}
\partial y_2^{j>}}  + \frac{4 G m_1}{c^4 r_1} v_1^{<i}
v_1^{j>}  +  \frac{4 G m_2}{c^4 r_2} v_2^{<i} v_2^{j>}.
\end{equation}
Consequently, the conformal factor, $\Psi$, must include a
perturbation, $\kappa $, to the ``Brill--Lindquist-like''
conformal factor, $\psi $, such that,
\begin{equation}
\label{perturbationconformalfactor} \Psi    =    \psi   + \kappa,
\end{equation}
where $\psi$ is given by (\ref{psi}) and its structure is chosen
to resemble as closely as possible the form of the time-symmetric,
Brill-Lindquist conformal factor, $\psi_{\mathrm{B-L}}$
(\ref{psiBL}).

We then assume  that $\kappa $, the perturbation in the conformal
factor, and $\widetilde{K}_{ij}$,  the full conformal extrinsic
curvature,  admit  power-like expansions of the type,
\begin{eqnarray}
\label{perturbationseries1} \kappa   &    =      &
\sum^{+\infty}_{n=2}  \left(  \frac{1}{c^{2n}}  \right)  \kappa_{  (n/2)  },\\
\nonumber\\
\label{perturbationseries2} \widetilde{K}_{ ij }   &    =    &
\sum^{ +\infty }_{ n=1 } \left( \frac{1}{c^{2n+1}} \right)
\widetilde{K}_{[n + 1/2]ij }.
\end{eqnarray}
Subtle considerations determine the form of the above expansions.
Notably, we introduce formally the expansions to investigate
perturbations to the Bowen-York solution \cite{BowY80}. They do
not, as they might otherwise misleadingly suggest, represent
post-Newtonian expansions in the general sense. Instead, the
parameter $c$ tracks the order of the perturbation to the
Bowen-York solution. Consequently, the lowest order of the
perturbation metric for a non-zero extrinsic curvature, $
h_{ij}|_{K_{ij} \neq 0} $, is to be of $ {\cal O} (1 /c^4) $. In
this light, it is therefore more appropriate to regard the
resulting perturbation equations (at low order) as successive
integer approximations in the perturbative metric, $h_{ij} $, of
the conformally-flat solution. Note that our definition
(\ref{perturbationseries2}) determines the zeroth order term in
$\widetilde{K}_{ij}$ (i.e. $\widetilde{K}_{[\frac{3}{2}]ij }$) to
be of ${\cal O} ( 1/c^3)$, which is dimensionally consistent with
both $K_{ij}^{\mathrm{2PN}}$ and $\tilde{K}_{ij}^{\mathrm{B-Y}}$,
as given later by (\ref{2PNExtCurv}) and
(\ref{simpledefinitionB-Y2}) respectively.

\subsubsection{Analytic closed form of the linearized solution}

By considering the formal perturbation, as defined by
(\ref{perturbationseries1}) and (\ref{perturbationseries2}), of
constraint (\ref{constraint3}) together with
(\ref{InitialConstraint1EC}) and (\ref{InitialConstraint1CTS}), we
arrive at the linearized perturbation of the Hamiltonian
constraint,
\begin{equation}
\label{linearperturbationB-Y} \widetilde{\Delta}  \kappa_{(1)}  =
h^{ij} \partial_{ij} \Psi  +  \partial_{i} h_{i}^{k}
\partial_{k} \Psi  +  \frac{1}{8} \Psi \partial_{ij} h_{ij}.
\end{equation}
The above is identical in form to its counterpart in the case of a
linear perturbation to the time-symmetric Brill-Lindquist
constraints, given by Eqn.~(3.7) in \cite{B03iv}. In contrast, the
linear--order terms in (\ref{linearperturbationB-Y}) in both the
perturbation metric, $h_{ij} $
(\ref{perturbationconformalflatness2}), and Brill-Lindquist-like
conformal factor, $\psi $ (\ref{psi}), include terms due to the
velocity of the black holes. On the other hand, we may identify
the time-independent parts in $h_{ij}$ and $\psi$,
(\ref{perturbationconformalflatness2}) and (\ref{psi})
respectively, and refer to the results given in Section III B in
\cite{B03iv} for solving the time-symmetric components in
(\ref{linearperturbationB-Y}). Substitution of
(\ref{perturbationconformalflatness2}) and
(\ref{perturbationconformalfactor}) into
(\ref{linearperturbationB-Y}), allows us to determine the exact
solution for $\kappa_{(1)}$,
\begin{eqnarray}
\nonumber \kappa_{(1)} & = & - G^2 m_1 m_2 \mathrm{D} \left[
\frac{g}{6} + \frac{r_1 + r_2}{4 r_{12}} \right] - \frac{G
m_1}{r_1} \left[ \frac{(n_1 v_1)^2}{4} - \frac{\mathbf{v_1}^2}{
12} \right] - \frac{G m_2}{r_2} \left[ \frac{(n_2 v_2)^2}{4} -
\frac{\mathbf{v_2}^2}{ 12} \right] \\ \nonumber & & + \alpha_1
\left\{ - G^2 m_1 m_2  \left[ 4 H_1 + \frac{K_1}{4} - \frac{1}{4}
\mathrm{D} \left( \frac{1}{r_2} \ln \left[ \frac{r_1}{r_{12}}
\right] \right) + \frac{9}{4} \mathrm{D} \left( \frac{\ln
r_1}{r_{12}} \right) + \frac{2 \mathrm{D}g}{3 r_1} \right. \right.
\\ \nonumber & & \left. \left. + 2 \Delta_1 \left(
\frac{g}{r_{12}} \right) - \frac{1}{4} \Delta_2 \left(
\frac{g}{r_{12}} \right) - \frac{1}{3 r_1 r_{12}^2} + \frac{1}{4
r_2 r_{12}^2} \right] + G m_1
  \left[ - \frac{35 (n_1 v_1)^2}{8 r_1^2} + \frac{35 \mathbf{v_1}^2}{24 r_1^2} \right] \right. \\
\label{kappa1HK} & & \left. + G m_2 \left[ 4 v_2^i v_2^j
\phantom{k}_{ij}g  +
  4 v_2^i v_2^j \phantom{k}_{(j} g_{i)} + \frac{v_2^i v_2^j}{2} g_{ij} -
  \frac{2 \mathbf{v_2}^2}{3 r_1 r_2} \right] \right\} + \alpha_2 \{ 1
  \leftrightarrow 2 \} \\ \nonumber & = & - G^2 m_1 m_2 \mathrm{D} \left[
\frac{g}{6} + \frac{r_1 + r_2}{4 r_{12}} \right] - \frac{G
m_1}{r_1} \left[ \frac{(n_1 v_1)^2}{4} - \frac{\mathbf{v_1}^2}{
12} \right] - \frac{G m_2}{r_2} \left[ \frac{(n_2 v_2)^2}{4} -
\frac{\mathbf{v_2}^2}{ 12} \right] \\ \nonumber & &
  + \alpha_1 \left\{ - G^2 m_1 m_2 \left[ 4
  \Delta_1 \left[ \frac{g}{r_{12}} + \mathrm{D} \left( \frac{r_1 + r_{12}}{2}
  g \right) \right] - 4 \mathrm{D} \left(\frac{\ln r_{12}}{r_1} \right) -
  \frac{15}{4} \mathrm{D} \left(\frac{\ln r_{1} }{r_{12}} \right) \right. \right. \\
  \nonumber & &  \left. \left. + \frac{2 \mathrm{D} g}{3 r_1} - \frac{1}{4} \Delta_2
  \left( \frac{g}{r_{12}} \right) - \frac{15 r_2}{8 r_1^2 r_{12}^2} +
  \frac{2}{r_1^2 r_{12}} - \frac{1}{8 r_1^2 r_{2}} - \frac{7 }{3 r_1 r_{12}^2}
  + \frac{3 }{8 r_2 r_{12}^2} \right] \right. \\
\nonumber  & & \left. + G m_1
  \left[ - \frac{35 (n_1 v_1)^2}{8 r_1^2} + \frac{35 \mathbf{v_1}^2}{24 r_1^2} \right]  + G m_2 \left[ 4 v_2^i v_2^j \phantom{k}_{ij}g  +
  4 v_2^i v_2^j \phantom{k}_{(j} g_{i)} + \frac{v_2^i v_2^j}{2} g_{ij} -
  \frac{2 \mathbf{v_2}^2}{3 r_1 r_2} \right] \right\} \\ \label{kappa1} & &
+  \alpha_2  \{  1  \leftrightarrow  2
 \},
\end{eqnarray}
where $\Delta_1 \equiv \frac{\partial^2}{\partial y_1^i
\partial y_1^i}$ and $\Delta_2 \equiv \frac{\partial^2}{\partial
y_2^i \partial y_2^i}$ denote the Laplacians with respect to the
source positions $\mathbf{y_1}$ and $\mathbf{y_2}$ respectively, and
$\mathrm{D} \equiv \frac{\partial^2}{\partial y_1^i
\partial y_2^i}$ as before. As pointed out earlier, the results (3.8) and (3.11) in \cite{B03iv} may be used in the above.  The expression for
$\kappa_{(1)}$ (\ref{kappa1}) is valid in the full distributional
sense, i.e. the addition of an arbitrary number of `homogeneous'
terms $\thicksim 1/r_1$ and $\thicksim 1/r_2$ is not permitted, as
mentioned earlier, when solving the Poisson-type equation for
$\kappa_{(1)}$. Note that $\kappa_{(1)}$ also tends to zero at
spatial infinity  (i.e. when $r \equiv |\mathbf{x}| \rightarrow +
\infty$). Consistent with Eqns.~(3.8) and (3.11) in \cite{B03iv},
only terms in the first line of (\ref{kappa1HK}) and (\ref{kappa1})
contribute at the 2PN order in the conformal factor, $\Psi$. The
additional terms, which are proportional to 1PN constants,
$\alpha_1$ or $\alpha_2$, appear only at the 3PN order. Indeed,
(\ref{kappa1}) reflects characteristic 3PN features, despite
imposing only the initial isometric relationship with a 2PN
expansion, (\ref{pNandconfmetricI}). In particular, the intermediate
expression, (\ref{kappa1HK}), contains the special Poisson-like
solutions $H_1$ and $K_1$, which appear in the 3PN spatial metric,
given by Eqn.~(111) in \cite{BFeom}\footnote{These occur in the
non-linear potential, $\hat{X}$.}. For completeness, we include here
the functions, $H_1$ and $K_1$, satisfying the following
Poisson-like equations,
\begin{eqnarray}
\label{poissonH1}
\Delta H_1  =  2 \phantom{k}_i g_{j} \partial_{ij} \left( \frac{1}{r_1} \right), \\
\label{poissonK1} \Delta K_1  =  2 \mathrm{D}^2  \left( \frac{\ln
r_1}{r_2} \right),
\end{eqnarray}
and can be found as,
\begin{eqnarray}
\nonumber
H_1 & = & \Delta_1  \left[  \frac{g}{2 r_{12}}  +  \mathrm{D}  \left( \frac{r_1 + r_{12}}{2} g \right) \right]  -  \mathrm{D} \left( \frac{\ln r_{12}}{r_1} \right)  -  \frac{3}{2} \mathrm{D} \left( \frac{\ln r_{1}}{r_{12}} \right)  -  \frac{r_2}{2 r_1^2 r_{12}^2} \\
\label{H1}
& &  +  \frac{1}{2 r_1^2 r_{12}}  -  \frac{1}{2 r_1 r_{12}^2}, \\
\label{K1}K_1 & = & \mathrm{D}  \left( \frac{1}{r_2} \ln \left[
\frac{r_1}{r_2} \right] \right)  -  \frac{1}{2 r_1^2 r_2}  +
\frac{1}{ 2 r_2 r_{12}^2}  +  \frac{r_2}{ 2 r_1^2 r_{12}^2}.
\end{eqnarray}
Finally, although (\ref{kappa1HK}) and (\ref{kappa1}) indicate 3PN
characteristics, it is important to note that they do not represent
complete expressions for $\kappa_{(1)}$ at 3PN. Such an expression
is only possible if we consider terms of 3PN order in $h_{ij}$ (i.e.
using 3PN results such as given in \cite{BFeom}), which we have not
attempted here. Furthermore, in order to be strictly correct up to
and including order ${\cal O} (1/c^6)$, we should instead solve for
$\kappa_{(3/2)}$ by considering the 1.5 linear order equation to the
conformally-flat Hamiltonian and momentum constraints, containing
the first explicit appearance of the extrinsic curvature, $K_{ij}$,
in the hierarchy or perturbative equations:
\begin{equation} \label{1.5pertHamconstraint} \widetilde{\Delta} \kappa_{(3/2) } -
h^{ij} \partial_{ij} \Psi  -  \partial_{i} h_{i}^{k}
\partial_{k} \Psi  -  \frac{1}{8} \Psi \partial_{ij} h_{ij} -
 \frac{1}{12} \Psi^5 K_{[3/2]}^2  +    \frac{1 }{8} \Psi^5 \widetilde{K}_{[3/2]ij}     \widetilde{K}^{ ij }_{[3/2]}
= 0, \end{equation}
\begin{equation} \label{1.5pertMomconstraint} \widetilde{\nabla}_{j}    \left(
\widetilde{K}^{ ij }_{[3/2]} - \delta^{ij} K_{[3/2]}   \right) =
 0. \end{equation}
Such an investigation is beyond the scope of the current work.

Finally, the fully explicit form of $\kappa_{(1)}$, obtained by
expanding all the derivatives in the result (\ref{kappa1}), is
given here for completeness as:
\begin{eqnarray} \nonumber \kappa_{(1)} & = &    -  G^2 m_1
m_2 \left[ \frac{1}{12 r_1 r_2} + \frac{r_1 + r_2}{8 r_{12}^3} +
\frac{1}{24 r_{12}} \left( \frac{1}{r_1} + \frac{1}{r_2} \right) -
\frac{1}{8 r_{12}^3} \left( \frac{r_1^2}{r_2} + \frac{r_2^2}{r_1}
\right) \right] \\ \nonumber & & - G m_1 \left[ \frac{(n_1
v_1)^2}{4 r_1} - \frac{\mathbf{v_1}^2}{12 r_1} \right] - G m_2
\left[ \frac{(n_2 v_2)^2}{4 r_2} - \frac{\mathbf{v_2}^2}{12 r_2}
\right] + \alpha_1 \left\{ - G^2 m_1 m_2 \left[ - \frac{2}{r_1^3}
- \frac{13}{8 r_{12}^3} \right. \right. \\ \nonumber & & \left.
\left. - \frac{1}{3 r_1 r_{12}^2} - \frac{5}{24 r_1^2 r_{12}} +
\frac{5}{24 r_1^2 r_2} - \frac{r_1}{4 r_2 r_{12}^3} + \frac{3}{8
r_2 r_{12}^2} - \frac{1}{3 r_1 r_2 r_{12}} + \frac{2 r_2}{r_1
r_{12}^3}  \right. \right. \\ \nonumber & & \left. \left.  -
\frac{15 r_2}{8 r_1^2 r_{12}^2} + \frac{2 r_2}{r_1^3 r_{12}} +
\frac{15 r_2^2}{8 r_1^2 r_{12}^3} + \frac{2 r_2^2}{ r_1^3
r_{12}^2} - \frac{2 r_2^3}{r_1^3 r_{12}^3} \right] + G m_1 \left[
- \frac{35 (n_1 v_1)^2 }{8 r_1^2} +
\frac{35 \mathbf{v_1}^2}{24 r_1^2} \right] \right. \\
\nonumber & & \left. + G m_2 \left[ - \frac{(n_{12} v_2)^2}{2 S^2} +
\frac{4 (n_{12} v_2) (n_1 v_2)}{S^2} - \frac{4 (n_{1} v_2)^2}{S^2} +
\frac{3 (n_{12} v_2) (n_2 v_2)}{S^2} - \frac{4 (n_{1} v_2) (n_2
v_2)}{S^2}  \right. \right. \\
\nonumber & & \left. \left.  - \frac{ (n_{2} v_2)^2}{2 S^2} - \frac{
4 (n_{1} v_2)^2}{r_1 S} - \frac{ (n_{12} v_2)^2}{2 r_{12} S} -
\frac{ (n_{2} v_2)^2}{2 r_{2} S} + \frac{ 4 \mathbf{v_2}^2}{r_{1} S}
+\frac{ \mathbf{v_2}^2}{2 r_{12} S} + \frac{ \mathbf{v_2}^2}{2 r_{2}
S} -
  \frac{2 \mathbf{v_2}^2}{3 r_1 r_2}
\right] \right\} \\ \label{kappaexplicit} & & + \alpha_2 \{ 1
\leftrightarrow 2 \}.
\end{eqnarray}
where $S = (r_1 +r_2 +r_{12})$.

\subsection{Choice of Extrinsic Curvature and Maximal Slicing}
\label{Extrinsic-Curvature-and-Maximal-Slicing}

Despite their manifestly different forms, we show here that the
physical extrinsic curvature of the conformally-flat Bowen-York
solution, $K_{ij}^{\mathrm{BY}}$, and the extrinsic curvature
derived from 2PN results, $K_{ij}^{\mathrm{ 2PN}}$, are physically
equivalent to each other modulo the infinitesimal coordinate
transformation specified by $\xi^{\mu}$, as given by (\ref{xi01})-
(\ref{xiiDirac}). In particular, we show the following relationship,
\begin{eqnarray}
K_{ij}^{\mathrm{BY}} & = & K^{\mathrm{2PN}}_{ij} + \partial_{ij} \xi_{0} +
{\cal O} \left(\frac{1}{c^{4}} \right), \label{changeofKij}
\end{eqnarray}
where $\xi_{0}$ refers to the $0$th (covariant) component of the
gauge vector $\xi^{\mu}$ and is specified by (\ref{xi01}). The
extrinsic curvature derived from post-Newtonian results
\cite{BFP98}, $K_{ij}^{\mathrm{ 2PN}}$, at 2PN order  is given by,
\begin{eqnarray}
\label{2PNExtCurv} K_{ij}^{\mathrm{2PN}} & = &
\frac{G m_1}{c^3 r_1^2} \left( 4 n_{1(i} v_{1j)} -
\delta_{ij}(n_1 v_1) \right) + 1 \leftrightarrow 2 + {\cal O}
\left(\frac{1}{c^5} \right),
\end{eqnarray}
On the other hand, the conformal Bowen-York extrinsic
curvature\footnote{Strictly speaking, the Bowen--York conformal
extrinsic curvature, $\tilde{K}_{ij}^{\mathrm{full}
\mathrm{B-Y}}$, is the solution to the constraints in the case of
conformally flat (i.e. $\tilde{\gamma_{ij}} = \delta_{ij}$) vacuum
spacetimes with a maximal hypersurface, $K=0$, and is explicitly
given by,
\begin{eqnarray}
\nonumber \tilde{K}_{ij}^{\mathrm{full} \mathrm{BY}} & = & \frac{3 G
m_1}{2 c^3 r_1^{2}} [ 2 v_{1(i} n_{1j)} - ( n_{1} v_{1} )
\left( \delta_{ij} - n_{1 i} n_{1 j} \right) ] \\
\label{definitionB-Y} &&  \mp  \frac{3 G a^2 m_1 }{2 c^3 r_1^4} [ 2
v_{(1i} n_{1j)}  +  ( n_{1} v_{1 } )  \left( \delta_{ij} - n_{1 i}
n_{1 j} \right) ]
 +  1  \leftrightarrow   2,
\end{eqnarray}
where we consider only the linear momentum term (with no intrinsic
angular momentum) of the original solution. The second term in
(\ref{definitionB-Y}) corresponds to its ``inversion-symmetric
term''; see \cite{BowY80}. This complete solution
(\ref{definitionB-Y}) was chosen historically to generalize the
two--sheeted topology of the Misner--Lindquist time--symmetric
approach. As a result, the inversion--symmetric term satisfies the
isometry condition for a field to exist on a two--sheeted
manifold. The constant $a$ in (\ref{definitionB-Y}) denotes the
radius of the inversion sphere, or alternatively, the throat of
the black hole after applying appropriate boundary conditions.
Importantly, our approach does not concern itself directly with
the topological nature of the two black holes. Therefore, we only
refer to part of the Bowen-York solution for the most general
topologies, $\tilde{K}_{ij}^{\mathrm{BY}}$ (henceforth referred to
as `Bowen-York'), where
\begin{eqnarray}
\label{simpledefinitionB-Y} \tilde{K}_{ij}^{\mathrm{BY}} & = &
\frac{3 G m_1}{2 c^3 r_1^{2}}  [ 2 v_{1(i} n_{1j)} -
 ( n_{1} v_{1})  \left( \delta_{ij} - n_{1 i} n_{1 j}
\right) ]  +  1  \leftrightarrow   2.
\end{eqnarray}By general topologies, we include both the two-sheeted
asymptotically flat universe Misner--Lindquist \cite{Misner63} and
three-sheeted Brill-Lindquist \cite{BrillLind63} solutions.},
$\tilde{K}_{ij}^{\mathrm{BY}}$, is given by,
\begin{eqnarray} \label{simpledefinitionB-Y2}
\tilde{K}_{ij}^{\mathrm{BY}} & = & \frac{3 G m_1}{2 c^3 r_1^{2}}  [
2 v_{1(i} n_{1j)} -
 ( n_{1} v_{1})  \left( \delta_{ij} - n_{1 i} n_{1 j}
\right) ]  +  1  \leftrightarrow   2.
\end{eqnarray}
We hence denote the physical Bowen-York extrinsic curvature as
$K_{ij}^{\mathrm{BY}} = \Psi^2 \tilde{K}_{ij}^{\mathrm{BY}}$, where
$\Psi$ is given by (\ref{Psi2pn}) in symmetric tracefree form and by
(\ref{psiDIRAC}) in Dirac coordinates. Therefore, by considering
terms up to ${\cal O} (1/c^5)$, $K_{ij}^{\mathrm{BY}} =
\tilde{K}_{ij}^{\mathrm{BY}} + {\cal O} (1/c^5)$.

Note that, at first sight, the relationship (\ref{changeofKij})
appears to contradict the standard transformation law of tensors.
For completeness, we state the standard transformation law of
tensors for $K_{ij}$ in the instance of an infinitesimal change of
coordinates, $\{x^{\mu}\} \rightarrow \{x'^{\mu}\} $,
$\bar{K}'_{\mu\nu}  = \frac{\partial x^{\sigma}}{\partial
x'^{\mu}} \frac{\partial x^{\rho}}{\partial x'^{\nu}}
K_{\sigma\rho}$,\private{
\begin{equation}
\label{kijstandardtransformation} \bar{K}'_{\mu\nu}  =
\frac{\partial x^{\sigma}}{\partial x'^{\mu}} \frac{\partial
x^{\rho}}{\partial x'^{\nu}} K_{\sigma\rho}.
\end{equation}}
 such that the spatial components of the extrinsic curvature
transform at linear order in $\xi^{\mu} $ are given by,
\begin{equation}
\label{kijstandardtransformation2} K_{ij}  = \bar{K}'_{ij} + 2
\partial_{(i} \xi_{j)} +  {\cal O} (\xi^{2}).
\end{equation}
If we relabel $K_{ij} \equiv K_{ij}^{\mathrm{2PN}}$, the
difference between (\ref{changeofKij}) and
(\ref{kijstandardtransformation2}) is immediately evident. We
stress, however, that despite this apparent contradiction, both
equations are valid. The distinctions between the two
3--dimensional extrinsic curvatures, $K^{\mathrm{B-Y}}_{ij} $
(which we relabel $K'_{ij}$ in this instance) and $\bar{K}'_{ij}$,
given by (\ref{changeofKij}) and
(\ref{kijstandardtransformation2}) respectively, reveal the nature
of the 3+1 foliation of the spacetime associated with a gauge
transformation\footnote{Specifically, the 3-extrinsic curvature,
$K'_{ij} $, given by the coordinate transformation
(\ref{changeofKij}), corresponds to a distinct re-formulation of
the 3--foliation of spacetime. That is the 3--dimensional
hypersurface $\Sigma'_t$, timelike unit vector $n'^{a} $, lapse
function $N' $, and vector shift $\beta'_{i} $, of the new
coordinate system, $\{x'^{\mu}\}$, are different physical entities
to the corresponding $\Sigma_t$, $n^{a} $, $N $ and $\beta_{i} $
of the original coordinate system, $\{x^{\mu} \} $.}.

Let us now outline the proof of the relationship
(\ref{changeofKij}) between the two extrinsic curvatures,
$K'_{ij}$ and $\bar{K}'_{\mu\nu}$. This can be understood by first
considering the infinitesimal difference, $\delta n'_{\nu}$,
between the time-like normal vectors, $\bar{n'}_{\mu}$ and
$n'_{\mu}$,
\begin{equation}
\label{deltan} \delta n'_{\nu} = \bar{n'}_{\nu} - n'_{\nu},
\end{equation}
where $\bar{n}'^{\mu}$ and $n'^{\mu}$ denote two distinct physical
vectors in the coordinate system $\{x'^{\mu}\} $, which result from
the transformation equations  (\ref{changeofKij}) and
(\ref{kijstandardtransformation2}) respectively. From $ n^{\mu}
\equiv - N \nabla^{\mu} t $ and using the standard tensor
transformation law (\ref{kijstandardtransformation2}), it is
possible to obtain explicit covariant and contravariant expressions
for (\ref{deltan}),
\begin{equation}
\label{deltanmucov2} \delta{n}'_{\nu} = \left(
\begin{array}{ccc}
N \partial_{0} \xi^{0} + N^{3} \partial_{0} \xi^{0} + {\cal O}(\xi^2) \\
N \partial_{i} \xi^{0}+ {\cal O}(\xi^2) \\
\end{array} \right),
\end{equation}
and
\begin{equation}
\label{deltanmucon2} \delta{n}'^{\mu}  =  \left(
\begin{array}{ccc}  -N \partial_{0} \xi^{0} -
\frac{\beta^{i} \partial_{i} \xi^{0}}{N}+ \frac{\partial_{0} \xi^{0}}{N}+
{\cal O}(\xi^2) \\ N \partial_{0} \xi^{i} + N \beta^{i} \partial_{0} \xi^{0}
+ N \partial_{i} \xi^{0} - \frac{\beta^{i} \partial_{k} \xi^{k}}{N} +
\frac{\partial_{0} \xi^{i}}{N} + {\cal O}(\xi^2) \\
\end{array} \right)
\end{equation}
respectively. If we then consider the quantity, $\delta
K'_{\mu\nu} $, the infinitesimal difference between $K'_{\mu\nu}$
and $\bar{K}'_{\mu\nu}$, i.e.
\begin{equation}
\label{changeofkij}
K'_{\mu\nu} = \bar{K}'_{\mu\nu} + \delta K'_{\mu\nu},
\end{equation}
where, following the definition (\ref{Kijdefn3}), we find $\delta
K'_{\mu\nu}$,
\begin{eqnarray}
\nonumber \delta K'_{\mu\nu} & = & \delta  \left( -
\gamma'^{\rho}_{(\mu} \nabla'_{\rho} n'_{\nu)} \right) \\
\label{deltakij}  & = & - \left(\nabla'_{(\mu} ( \delta n'_{\nu)} )  +  (
\delta n'^{\rho} ) n'_{(\mu} \nabla'_{\rho} n'_{\nu)}  +  (
n'^{\rho} ) ( \delta n'_{(\mu} ) \nabla'_{\rho} n'_{\nu)} \right).
\end{eqnarray}
Substituting (\ref{deltanmucov2}) and (\ref{deltanmucon2}) in
(\ref{deltakij}) and using $\gamma_{\mu \nu} = g_{\mu \nu} +
n_{\mu} n_{\nu}$, the hypersurface projection of the change in
extrinsic curvature, $\delta K'_{ij} $, is given as,
\begin{eqnarray}
\nonumber \delta K'_{ij} & = & - \frac{\partial ( \delta n'_{j)} )}{\partial
x'^{(i}} +\phantom{i}^{(4)}\Gamma'^0_{ij} ( \delta n'_{0} ) +
\phantom{i}^{(4)}\Gamma'^k_{ij} ( \delta n'_{k} ) + \delta n'^{0} n'_{(i}
\phantom{i}^{(4)}\Gamma'^0_{0j)} n'_{0} + \delta n'^{k} n'_{(i}
\phantom{i}^{(4)}\Gamma'^0_{kj)} n'_{0} {} \\ \nonumber & &{}+ n'^{0} \delta
n'_{(i} \phantom{i}^{(4)}\Gamma'^0_{0j)} n'_{0} + n'^{k} \delta n'_{(i}
\phantom{i}^{(4)}\Gamma'^{0}_{kj)} n'_{0},
\end{eqnarray}
where $\phantom{i}^{(4)}\Gamma'^{\mu}_{\nu \rho}$ refers to the 4-dimensional
Christoffel symbol. Finally, using $n'_{\mu} = (-N',0)$ and $n'^{\mu}= 1/N' (1
, -\beta'^{i})$, we obtain at linear order in $\xi^{\mu}$,
\begin{eqnarray}
\nonumber \delta K'_{ij} & = & - N \partial_{ij} \xi^{0} - \frac{\partial_{0}
\xi_{0} ( 1 + N^{2} )}{2 N} ( \partial_{j} \beta_{i} + \partial_{i} \beta_{j}
- \partial_{0} \gamma_{ij} ) + \frac{\partial_{0} \xi_{0} ( 1 + N^{2} )
\beta^k}{2 N} ( \partial_{j} \gamma_{ki} + \partial_{i} \gamma_{kj} -
\partial_{k} \gamma_{ij} ) {} \\ \nonumber & &{}+ \frac{\beta^{k} \partial_{k}
\xi^{0}}{2 N} ( \partial_{j} \beta_{i} + \partial_{i} \beta_{j} - \partial_{0}
\gamma_{ij} ) + \frac{N \gamma^{km} \partial_{k} \xi^{0}}{2} ( \partial_{j}
\gamma_{mi} + \partial_{i} \gamma_{mj} - \partial_{m} \gamma_{ij} ) {} \\
\nonumber & &{}- \frac{\beta^{k} \beta^{m} \partial_{k} \xi^{0}}{2 N}
(\partial_{j} \gamma_{mi} + \partial_{i} \gamma_{mj} - \partial_{m}
\gamma_{ij} ) + \frac{\partial_{(i} \xi^{0}}{2N} ( \partial_{j)} ( - N^{2} +
\beta_{k} \beta^{k} )) {} \\ \label{finaldeltakijlast} & &{}- \frac{\beta^{k}
\partial_{(i} \xi^{0}}{N} ( \partial_{j)} \beta_{k} ) + \frac{\beta^{k}
\beta^{m} \partial_{(i} \xi^{0}}{2 N} ( \partial_{j)} \gamma_{mk} +
\partial_{k} \gamma_{mj)} - \partial_{m} \gamma_{kj)} ) + {\cal O} ( \xi^{2}
).
\end{eqnarray}
By considering terms up to and including ${\cal O}
(\frac{1}{c^3})$, (\ref{finaldeltakijlast}) simplifies to the
recognizable form of the relationship (\ref{changeofKij}),
\begin{equation}
\label{deltakijc3} \delta K'_{ij}  =    \partial_{ij} \xi_{0} +
{\cal O}  \left(\frac{1}{c^{4}}\right).\end{equation}

Having proved the physical equivalence between the 2PN derived
$K_{ij}^{\mathrm{2PN}}$ and the standard Bowen-York
$K_{ij}^{\mathrm{BY}}$, it is then possible to use either the free
data $(K_{ij}^{\mathrm{2PN}}, K^{\mathrm{2PN}})$ or
$(K_{ij}^{\mathrm{BY}}, K\equiv 0)$. These results were given in
both EC and CTS decompositions in Sections~\ref{EC-Decomposition}
and \ref{CTS-Decomposition} respectively.

We finally provide complementary results to the PN free data
presented in Sections~\ref{EC-Decomposition} and
\ref{CTS-Decomposition}. Specifically, by assuming the widely used
boundary conditions at spatial infinity, $\Psi|_{r \rightarrow
\infty} = 1$ and $X^{i}|_{r \rightarrow \infty} = 0$, we solve for
the constrained variables, $X^{i}$ and $\beta^i$, to 2PN order in
the EC and CTS decompositions respectively:

\vspace*{0.1in}

\noindent {\em 1. EC decomposition}
%\subsubsection{EC decomposition}
\begin{enumerate}

\item[(a)] Maximal hypersurface, $K=0$

\begin{enumerate}

\item[i)] $\tilde{\sigma} = 2 \tilde{N}$
\begin{equation}
\label{Xtildei2} X^{i}  =  -\frac{ G m_1}{2
c^3 r_1}  \left( 7 v_1^i + n_1^i (n_1 v_1) \right)   +  1 \leftrightarrow 2 +  {\cal O} \left( \frac{1}{c^5} \right).
\end{equation}

\item[ii)] $\tilde{\sigma}^{\mathrm{CTT}}$

\begin{equation}
\label{Xtildei2} X^{i}  =  - \frac{ G m_1}{4
c^3 r_1}  \left( 7 v_1^i + n_1^i (n_1 v_1) \right)+  1 \leftrightarrow 2    +  {\cal O} \left( \frac{1}{c^5} \right).
\end{equation}

\end{enumerate}

\item[(b)] Mean curvature of $K^{\mathrm{2PN}}$

\begin{enumerate}

\item[i)] $\tilde{\sigma} = 2 \tilde{N}$
\begin{equation}
\label{Xtildei} X^{i}  =   -  \frac{4 G m_1}{c^3 r_1} v_1^i   +  1 \leftrightarrow 2  +  {\cal O} \left( \frac{1}{c^5} \right).
\end{equation}
\item[ii)] $\tilde{\sigma}^{\mathrm{CTT}}$

\begin{equation}
\label{Xtildei} X^{i}  =   -  \frac{2 G
m_1}{c^3 r_1} v_1^i +  1 \leftrightarrow 2  + {\cal O} \left( \frac{1}{c^5} \right).
\end{equation}

\end{enumerate}

\end{enumerate}

\vspace*{0.1in}

\noindent {\em 2. CTS decomposition}
%\subsubsection{CTS decomposition}

\begin{enumerate}

\item[(a)] Maximal hypersurface, $K=0$

\begin{equation}
\label{betai} \beta^{i} = -  \frac{G m_1}{2 c^3 r_1}  \left( 7
v_{1}^i + n_{1}^i (n_1 v_1) \right) + 1 \leftrightarrow 2 + {\cal O} \left(
\frac{1}{c^5} \right).
\end{equation}
\item [(b)] Mean curvature with $K^{\mathrm{2PN}}$

\begin{equation}
\label{betai} \beta^{i} = -  \frac{4 G m_1}{ c^3 r_1} v_{1}^i  +  1 \leftrightarrow 2  +
{\cal O} \left(\frac{1}{c^5} \right).
\end{equation}

\end{enumerate}

\section{Conclusion}
\label{conclusion}

This work provides astrophysically realistic free data for binary
black holes in numerical relativity, which are in agreement with
2PN results. Following the time-symmetric approach of
\cite{B03iv}, we propose a particular solution to the constraint
equations in the form of the standard conformal decomposition of
the spatial metric and extrinsic curvature. The solution presented
here is shown to differ from the post-Newtonian metric in harmonic
coordinates up to 2PN order by a coordinate transformation. The
solution is also shown to differ at 2PN from the conformally-flat
Bowen-York solution of the constraints, despite the singular
nature of the proposed conformal metrics, $\tilde{\gamma}_{ij}$
and $\tilde{\gamma}_{ij}^{\sl{Dirac}}$. We recall that the
post-Newtonian metric is not only valid in the `near-zone' of
BBHs, but also arises from the re-expansion of a `global'
post-Minkowskian multipole expansion when $c \rightarrow \infty$,
which is equivalent to a far-zone expansion when $r \rightarrow
\infty$. In addition, the post-Newtonian masses $m_1$ and $m_2$
are introduced as coefficients of Dirac delta functions in the
Newtonian density of point-like particles. Together with the
formal energy and mass calculations for the time-symmetric
instance \cite{B03iv}, the interpretation of our solution as a
perturbation of the Bowen-York solution suggests that the
post-Newtonian description of the black holes as delta-function
singularities agrees with the physical masses of the Bowen-York
black holes. The latter are computed by surface integrals at
infinity and associated with Einstein-Rosen-like bridges.

We note, however, that our solution does not include the 2.5PN term
of the metric, associated with Newtonian radiation reaction effects.
Furthermore, although our solution exhibits characteristic 3PN
features, we do not directly use the considerably more complex 3PN
metric itself. In addition, we have only considered systems of two
non-spinning black holes. However, spin effects are known to
contribute directly to the gravitational waveform, and to the
overall emission of energy and angular momentum of the system
\cite{KWWi93,K95,0TO98,TOO01}. Finally, it is important to note that
further studies are required on the behavior of our solution in the
vicinity of the black holes, providing an insight into the precise
nature of the singularity, which is necessary for practical
numerical implementation of the data.

\acknowledgments I wish to thank my supervisor, Luc Blanchet, for
suggesting this project and for valuable comments on this
research. In addition, I am grateful to Luca Baiotti, Bernard
Kelly, Richard Matzner, Harald Pfeiffer, Mark Scheel, Koji Uryu
and Bernard Whiting for valuable comments on the manuscript and
suggestions. I also wish to thank Silvano Bonazzola, Erin Bonning,
Manuela Campanelli, Sergio Dain, Eric Gourgoulhon, Philippe
Grandcl{\'e}ment and Carlos Lousto, Jose Luis Jaramillo Martin and
Frans Pretorius for their comments and useful discussions.
Finally, I wish to express my gratitude to the anonymous referee
for his/her comments; these have resulted in significant
improvements in the manuscript.

I gratefully acknowledge the Leverhulme Trust `Study Abroad' and Entente
Cordiale Scholarships and Robert Blair Fellowship (Corporation of London) for
financial support during the course of this work.

\appendix

\section{Comparison with the Kerr-Schild initial data}
\label{Kerr-Schild-initial-data}

One of the most widely--used set of initial data, which assumes a
deviation from a conformally flat spacetime, is based on the
superposition of two Kerr black holes in Kerr-Schild coordinates
\cite{MHS99,MaM00,BoMaNM03}. When comparing with PN calculations,
we find, however, that the `physically realistic' 2PN spatial
metric, $g_{ij}^{\mathrm{2PN}}$, disagrees with the free data
constructed by superposing two Kerr-Schild metrics together. In
particular, we show that the physical metric,
$\gamma_{ij}^{\mathrm{Kerr-Schild}}$, generated from a
numerically-computed conformal factor,
$\Psi^{\mathrm{Kerr-Shild}}$, is inconsistent with post-Newtonian
calculations at the 2PN level.

The Kerr-Shild spacetime metric for a single black hole of mass, $m$, and
specific angular momentum, $a = j/m$ (where $j$ is the black hole's angular
momentum), is given by,
\begin{equation}
\label{kerr-shildmetric} ds^2|_{\mathrm{Kerr-Schild}}  = \eta_{\mu
\nu} dx^{\mu} dx^{\nu}  +  2 H (x^{\alpha}) l_{\mu} l_{\nu}
dx^{\mu} dx^{\nu},
\end{equation}
where $\eta_{\mu \nu}$ is the Minkowski flat-space metric, $H
(x^{\mu})$ represents a scalar function, and $l_{\mu}$ denotes the
ingoing null vector (with respect to both the background and full
metric) such that $\eta_{\mu \nu} l_{\mu} l_{\nu}  = g^{\mu \nu}
l_{\mu} l_{\nu}  =  0$ (and hence, $l_0^2 = l_i l_i$). For a
general Kerr-Schild black hole metric (expressed in Kerr's
original rectangular coordinates), $H (x^{\mu})$ and $l_{\mu}$
are,
\begin{equation}
\label{kerr-shildmetricH} H  =  \frac{mr^3}{r^4 +
(\mathbf{a}.\mathbf{x})^2},
\end{equation}
and
\begin{equation}
\label{kerr-shildmetriclmu} l_{\mu}  =  \left(  1,  \frac{ r
\mathbf{x}  -  \mathbf{a} \times \mathbf{x}  +  (\mathbf{a} .
\mathbf{x})  \mathbf{a}/r}{r^2 + a^2}  \right),
\end{equation}
where $r$ is given by,
\begin{equation}
\label{kerr-shildmetriccoordcond2} r^2  =  \frac{1}{2}  \left(
\mathbf{x}^2 - \mathbf{a}^2 \right)   +  \sqrt{ \frac{1}{4}
(\mathbf{x}^2 - \mathbf{a}^2)^2 + (\mathbf{a}.\mathbf{x})^2}.
\end{equation}

From the spacetime metric (\ref{kerr-shildmetric}), we obtain the
3-spatial metric, $\gamma_{ij}$, within the ADM decomposition,
\begin{equation} \label{kerr-shildmetric-3spatial}
\gamma_{ij}^{\mathrm{Kerr-Schild}}  =  \delta_{ij}  +  2 H l_i
l_j,
\end{equation}
together with the ADM gauge variables $\beta_i = 2 H l_0 l_i$ and
$\alpha=\frac{1}{\sqrt{1+ 2 H l_0^2}}$. From (\ref{evolution1}),
the extrinsic curvature $K_{ij}^{\mathrm{Kerr-Schild}}$ is thus
given as,
\begin{equation} \label{extrinsiccurvatureADM}
K^{\mathrm{Kerr-Schild}}_{ij}  =  \frac{1}{2 \alpha} \left[
\nabla_j \beta_i + \nabla_i \beta_j - \partial_t
\gamma_{ij}^{\mathrm{Kerr-Schild}} \right].
\end{equation}

The Kerr-Schild conformal metric of two black holes,
$\tilde{\gamma}_{ij}^{\mathrm{Kerr-Schild}}$, is generated from
the `superposition' of two Kerr-Schild coordinate systems
(\ref{kerr-shildmetric-3spatial}), each describing a single black
hole (see Eqn.~(28) in \cite{MHS99} and Eqn.~(21) in
\cite{BoMaNM03}), i.e.
\begin{equation}
\label{kerr-shildmetric-3spatialconformal}
\tilde{\gamma}_{ij}^{\mathrm{Kerr-Schild}}  =  \delta_{ij}  +  2
\phantom{i}_1 H (r_1) \phantom{i}_1 l_i \phantom{i}_1 l_j +  2
\phantom{i}_2 H (r_2) \phantom{i}_2 l_i \phantom{i}_2 l_j,
\end{equation}
where the indices 1 and 2 label the two black holes. Similarly,
the mean curvature is given as,
\begin{equation}
\label{kerr-shildmeancurvature} K^{\mathrm{Kerr-Schild}} =
\phantom{i}_1 (K_i^i)^{\mathrm{Kerr-Schild}} + \phantom{i}_2
(K_i^i)^{\mathrm{Kerr-Schild}},
\end{equation}
and the conformal symmetric trace-free extrinsic curvature is,
\begin{eqnarray}
\nonumber (\tilde{A}_{ij})^{\mathrm{Kerr-Schild}} & =&
(\tilde{\gamma}_{k(i})^{\mathrm{Kerr-Schild}} \left[ \phantom{i}_1
( K_{j)}\phantom{i}^k)^{\mathrm{Kerr-Schild}} -\frac{1}{3}
\delta_{j)}\phantom{i}^k \phantom{i}_1
(K_i\phantom{i}^i)^{\mathrm{Kerr-Schild}} \right. \\
\label{kerr-shildAijtilde}  & & \left. + \phantom{i}_2
(K_{j)}\phantom{i}^k)^{\mathrm{Kerr-Schild}} -\frac{1}{3}
\delta_{j)}\phantom{i}^k \phantom{i}_2
(K_i\phantom{i}^i)^{\mathrm{Kerr-Schild}} \right]
\end{eqnarray}
where $\phantom{i}_1 (K_{i}^{k})^{\mathrm{Kerr-Schild}} =
\frac{1}{2 \alpha} \phantom{i}_1
(\gamma^{kj})^{\mathrm{Kerr-Schild}} \left[ \nabla_j \phantom{i}_1
\beta_i + \nabla_i \phantom{i}_1 \beta_j -
\partial_t \phantom{i}_1 (\gamma_{ij})^{\mathrm{Kerr-Schild}}
\right]$ and $\phantom{i}_1 (\gamma^{ij})^{\mathrm{Kerr-Schild}} =
2 \phantom{i}_1 H (r_1) \phantom{i}_1 l_i \phantom{i}_1 l_j$; see
\cite{BoMaNM03} for details. Note that for simplicity, we have not
included the `attenuation functions' $\phantom{i}_1 B$ and
$\phantom{i}_2 B$, introduced in \cite{MaM00,BoMaNM03}.

For illustrative purposes, we consider here the simplest instance
of two non-spinning black holes (i.e. $a = 0$), where the
Kerr-Schild 3-conformal metric is given explicitly as,
\begin{equation}
\label{kerr-shildmetric-3spatial2}
\tilde{\gamma}_{ij}^{\mathrm{Kerr-Schild}}   = \delta_{ij}
 +  \frac{2 G m_1}{c^2 r_1} n_1^i n_1^j  +  \frac{2 G
m_2}{c^2 r_2} n_2^i n_2^j,
\end{equation}
and the Hamiltonian constraint (\ref{constraint3}) is of the form,
\begin{eqnarray}
\nonumber \tilde{\Delta} \Psi^{\mathrm{Kerr-Schild}}  & =&
\frac{1}{8} ( \tilde{R} \Psi^{\mathrm{Kerr-Schild}}  )+
 \frac{(\Psi^{\mathrm{Kerr-Schild}})^5}{12}
 (K^{\mathrm{Kerr-Schild}})^2 \\
& & \label{HamiltonianconstraintTSym}
 -    \frac{(\Psi^{\mathrm{Kerr-Schild}})^{-7}}{8}  (\tilde{A}_{ij})^{\mathrm{Kerr-Schild}} (\tilde{A}^{ij})^{\mathrm{Kerr-Schild}}.
\end{eqnarray}
where $K^{\mathrm{Kerr-Schild}}$ and
$\tilde{A}_{ij}^{\mathrm{Kerr-Schild}}$ are of order ${\cal O}
(1/c^3)$ using (\ref{kerr-shildmeancurvature}) and
(\ref{kerr-shildAijtilde}).

Let us now assume that the Kerr-Schild conformal factor,
$\Psi^{\mathrm{Kerr-Schild}}$, at the 1PN order, is given by,
\begin{equation}
\label{PsiKerr-SchildTSym} \Psi^{\mathrm{Kerr-Schild}}  =
 1  +  \frac{Y}{c^2}  +  \left( \frac{1}{c^4} \right),
\end{equation}
which results in an explicit expression for the function $Y$ from
(\ref{HamiltonianconstraintTSym}),
\begin{equation}
\label{YKerr-Schild} Y  =  -  \frac{G m_1}{2 r_1}  - \frac{G
m_2}{2 r_2} + \frac{A G m_1}{r_1} + \frac{B G m_2}{r_2}.
\end{equation}
where $\sim \frac{A G m_1}{r_1}$ and $\sim \frac{B G m_2}{r_2}$
are additional `homogeneous' terms, which occur when solving
(\ref{PsiKerr-SchildTSym}). Note that when
$\Psi^{\mathrm{Kerr-Schild}} \equiv \psi^\mathrm{B-L}$, the
constants $A = B = 1$, where $\psi^\mathrm{B-L}$ is the
Brill-Lindquist conformal factor given by (\ref{psiBL}).

By substituting (\ref{PsiKerr-SchildTSym}) and
(\ref{YKerr-Schild}) into the conformal relationship,
$\gamma_{ij}^{\mathrm{Kerr-Schild}} =
(\Psi^{\mathrm{Kerr-Schild}})^4
\tilde{\gamma}_{ij}^{\mathrm{Kerr-Schild}}$, we find that
constants $A=B= 1/2$ in (\ref{YKerr-Schild}) for the following
isometric relationship to be true {\sl at 1PN order},
\begin{equation}
\label{KSPNequiv} \gamma_{ij}^{\mathrm{Kerr-Schild}} =
g_{ij}^{\mathrm{1PN}} + \delta_i \varsigma_j + \delta_j
\varsigma_i,
\end{equation}
where $g_{ij}^{\mathrm{1PN}}$ is the 1PN spatial metric in
harmonic coordinates given by Eqn.~(7.2c) in \cite{BFP98}. The
infinitesimal spatial gauge vector, $\varsigma_i$, is uniquely
determined as,
\begin{equation}
\label{varsigma} \varsigma_i = - \frac{G m_1}{c^2} n_1^i  -
\frac{G m_2}{c^2} n_2^i + {\cal O} \left( \frac{1}{c^4} \right).
\end{equation}
Using $A=B= 1/2$ in (\ref{YKerr-Schild}) and inserting
$\Psi^{\mathrm{Kerr-Schild}}$ into (\ref{conformalmetricdefn}), it
is immediately apparent that the physical metric,
$\gamma_{ij}^{\mathrm{Kerr-Schild}}$, is {\sl not} isometric with
$g_{ij}^{\mathrm{2PN}}$ - the 2PN spatial metric in harmonic
coordinates given by Eqn.~(7.2c) in \cite{BFP98}. More
specifically, $\tilde{\gamma}_{ij}^{\mathrm{Kerr-Schild}}$
(\ref{kerr-shildmetric-3spatialconformal}), does not contain any
`interaction' terms $\sim m_1 m_2$ of the two black holes.
Notably, there are no terms involving the `interaction' function
$g$ (\ref{g}), where we recall that $g$ and its associated
derivatives (such as (\ref{Dg}) and (\ref{dijigj})) are
characteristic of post-Newtonian results at 2PN order and higher.
The absence of such terms in
$\tilde{\gamma}_{ij}^{\mathrm{Kerr-Schild}}$ is a direct
consequence from its construction as a {\sl linear} superposition
of two Kerr-Schild coordinate systems. Note that it is impossible
to incorporate all the 2PN features of $g_{ij}^{\mathrm{2PN}}$
into higher order terms of the conformal factor
$\Psi^{\mathrm{Kerr-Schild}}$ (\ref{PsiKerr-SchildTSym}).

We, therefore, conclude that the Kerr-Schild conformal metric is
incompatible with the inspiral physics described accurately by 2PN
results.

  \bibliography{N06new}

\end{document}